\documentclass[prb,twocolumn,showpacs,showkeys,superscriptaddress]{revtex4}
\usepackage{amsmath}
\usepackage{amsfonts}
\usepackage{amssymb}
\usepackage{graphicx}
\usepackage{epstopdf}% for eps to pdf conversion
\usepackage{float} % for floating figures
\usepackage[caption=false]{subfig}
\usepackage[usenames,dvipsnames,svgnames,table]{xcolor} % for coloring
\usepackage[normalem]{ulem} % for crossing out text
\usepackage[makeroom]{cancel} % package for cancelling out in formulas
\DeclareMathOperator{\arcsinh}{arcsinh}

\begin{document}
\title{\textbf{Phase transitions in ensembles of solitons induced by an
optical pumping or a strong electric field.}}
\author{P. Karpov}
\affiliation{National University of Science and Technology "MISiS", Moscow, 119049, Russia}
\author{S. Brazovskii}
%\affiliation{National University of Science and Technology "MISiS", Moscow, 119049, Russia}
\affiliation{CNRS UMR 8626 LPTMS, University of Paris-Sud, University of Paris-Saclay, Orsay, France}

\begin{abstract}
The latest trend in studies of  modern electronically and/or optically active materials is to provoke phase transformations induced by high electric fields or by short (femtosecond) powerful optical pulses. The systems of choice are cooperative electronic states whose broken symmetries give rise to topological defects. For typical quasi-one-dimensional architectures, those are the microscopic solitons taking from electrons the major roles as carriers of charge or spin.
Because of the long-range ordering, the solitons experience unusual super-long-range forces leading to a sequence of phase transitions in their ensembles: the higher-temperature transition of the confinement and the lower one of aggregation into macroscopic walls.
Here we present results of an extensive numerical modeling for ensembles of both neutral and charged solitons in both two- and three-dimensional systems.
We suggest a specific Monte Carlo algorithm preserving the number of solitons, which substantially facilitates the calculations, allows to extend them to the three-dimensional case and to include the important long-range Coulomb interactions. The results confirm the first confinement transition, except for a very strong Coulomb repulsion, and demonstrate a pattern formation at the second transition of aggregation.
\end{abstract}

\date{August 2, 2016}
\maketitle

\section{Introduction}

\subsection{Solitons via doping or pumping.}

A new trend in controlling cooperative states of interacting electronic
systems is applying either very short (tens of femtoseconds) powerful optical
pulses or very high (up to $10^{7}$ V/cm) electric fields ("the electrostatic doping"), see Refs.
[\onlinecite{PIPT-book:2004,PIPT:04,PIPT5,FET,impact:12,impact:16}]. These impacts result in a very high (up to 10\% per lattice site) concentration of
excitations or charge carriers.
There are convincing arguments that these states will not resemble ensembles of
electrons and/or holes like for optical pumping or field-effect injection in
conventional semiconductors and can include superconductivity, antiferromagnetism, ferroelectricity, charge order, charge- and spin-density waves, Mott and Peierls insulators. The reason is that the commonly exploited
strongly correlated electronic systems show various types of symmetry breaking
giving rise to degenerate ground states. The degeneracy allows for
topologically nontrivial configurations exploring the possibility of traveling
through different allowed ground states. Their most known forms are plain
domain walls, stripes, vortex lines, or dislocations, which are still macroscopic
objects extending in one or two dimensions. Most importantly, there are also
totally localized and truly microscopic objects whose energies and quantum
numbers are on the one-electron scale. These anomalous particles -- the
solitons -- can determine the observable properties, which are usually ascribed
to conventional electronic excitations, see  Refs.
[\onlinecite{BK:84,YuLu}] for early theory reviews and  Refs.
[\onlinecite{SB-JSNM:2007,SB-Landau100:2009,JSSS:08}] for
updates.

The fact that the solitons have quantum eigenvalues (charge or spin) makes it
possible to control and monitor their concentration. The electrostatic doping (see the review \cite{FET} and updates in \cite{impact:12,impact:16})
should give rise to a stable $2D$ ensemble of similarly
charged kinks in a thin, sometimes atomically narrow, surface layer. The optical
pumping should give rise firstly to an equal number of oppositely charged
solitons, whose collisions will work to convert them secondly to an ensemble
of neutral spin-carrying solitons (which usually have lower energy than the
charged ones \cite{vardeny}). Inevitably, recombination will follow, possibly via the formation
of excitons as pairs of oppositely charged kinks \cite{BK:81}. However, the optical
emission will take long (more than nanoseconds) time, which can be further
prolonged by the intermediate conversion to neutral spin-carrying solitons,
which can recombine only via triplet channels -- the effect is well documented in the
optics of conducting polymers \cite{vardeny,spin-kinks}. Then, for a typically very
fast pump-induced phase transition experiment, even the system of oppositely charged solitons, and even more of spin-carrying ones, can be
treated as quasi stationary, with only slowly decreasing number of particles. This number is exactly conserved and monitored in experiments with
the electrostatic doping.

Such ensembles of solitons are expected to have a peculiar phase diagram, with several lines of phase transitions, which are inevitably crossed in
the course of the evolution or monitoring the concentration and the temperature. The study of these transitions is the main goal of this paper. In
the next section of Introduction, we shall summarize modern experimental and theoretical evidences in favor of existence of solitons in
electronic systems. In Sec.II, we shall give a qualitative picture of phase transitions in ensembles of neutral and charged solitons. The central
issue is the confinement interaction specific to solitons in contrast to conventional electrons. In Sec.III, we shall introduce the basic model,
which is mapped onto the constrained Ising model, and also present some analytical results. In Sec.IV, we shall present the results of extensive numerical
modeling, which was challenging for three-dimensional systems, particularly in presence of Coulomb interactions (CI). We shall observe a high-temperature transition of confinement of solitons into pairs and a particularly rich and interesting low-temperature transition of soliton
aggregation into domain walls transversing the sample. Sec.V is devoted to discussion and summary. Appendices contain estimations for the aggregation transition for neutral and charged solitons.

\subsection{New accesses to microscopic solitons in quasi-one-dimensional electronic systems.}
\label{New_accesses_to microscopic_solitons_in_q1D_electronic_systems}

There are growing experimental evidences on existence of microscopic solitons
and their determining role in electronic processes of quasi-1D systems:
conjugated polymers (see \cite{YuLu} on theory and \cite{Heeger RMP} on
experiment), spin-Peierls chains \cite{horvatic}, donor-acceptor stacks
including "electronic ferroelectrics" \cite{NI-solitons,Okamoto-NI}, and families of the so-called "electronic crystals", particularly charge density waves (CDW)
and charge-ordered Mott insulators (see reviews \cite{SB-Landau100:2009,SB-JSNM:2007,sb-cofe}). Solitons take over band electrons in roles of
primary excitations -- charge or spin carriers, since their activation energies are typically lower than gaps opened in the electronic spectra.
The solitons
feature self-trapping of electrons into mid-gap states and separation of spin
and charge into spinons and holons, sometimes with their reconfinement at
essentially different scales. Thus, the ferroelectric charge ordering in
organic conductors (see a review \cite{sb-cofe}) gives access to several types
of solitons observed in conductivity (holons) and in permittivity (polar
kinks), to soliton bound pairs in optics, to compound charge-spin solitons.
In CDWs \cite{SB-PRL:2012} and in surface nano-wires \cite{Chao-NC:2016} the
individual solitons, which are the amplitude kinks, have been visually
captured in STM experiments; this is the most remarkable new achievement in
proving the existence of solitons. The resolved subgap tunneling spectra
\cite{lat:05} recover presumably the same solitons in dynamics. The tunneling creation of
soliton pairs describes nonlinear transport in CDWs \cite{Miller} and
polymers \cite{park,NK+SB on Park}. The solitons can be also viewed as
nucleus of the melted stripe phase in doped Mott insulators or of the FFLO
phase in spin polarized superconductors \cite{buzdin,buzdin-machida}.

On this basis one can extrapolate to a picture of combined topological
excitations in general strongly correlated systems: from doped
antiferromagnets to strong-coupling and spin-polarized superconductors
\cite{SB-Landau100:2009}.

\section{Interactions and phase transitions in ensembles of solitons.}

\subsection{Interactions of solitons}

Following the above quoted experimental confirmations and theoretical results,
we shall consider a system where the solitons serve as the lowest (with
respect to band electrons) energy forms of storage of the charge or the spin.
We shall restrict the study to the case of a discrete symmetry breaking, which
is a very common phenomenon of a dimerization of bonds (the family of
Peierls-like transitions) or of sites (the family of transitions with charge
ordering or disproportionation).
Such a system possesses typically three types of
solitons: spinless ones with charges $\pm e$ and a neutral one with the spin $1/2$. In
ferroelectric systems (see a recent review \cite{NK+SB-FE:2016}) all solitons
should carry noninteger charges. With passing of such a soliton, the order parameter changes the sign, hence the nickname "the amplitude kink",
or just the "kink" or the soliton, which we shall use in the following.
Cases of continuous symmetries like
incommensurate CDWs, spin density waves and superconductors, require
special consideration.

The solitons are subject to all kinds of interactions, among themselves and
also with the lattice, known for conventional electrons.
The important CI can be screened or not screened by external carriers and we shall consider both cases.
But beyond that, there is the unusual super long-range interaction specific to the
solitons as topologically nontrivial objects. The soliton is a $1D$ domain boundary that interrupts the proper
interchain arrangement. That gives rise to the confinement energy $Fl$, with
the constant confinement force $F$, growing linearly with the distance between solitons $l$
-- see Fig.\ref{fig:T>T1}. This energy dominates at long
distances even if it can be unimportant locally for a crystal of weakly
interacting chains.

In some special cases, the ground-state degeneracy can be lifted by an
internal effect globally -- for the whole system. A bright example is the
cis-polyacetylene \cite{BK:81,BK:84,Heeger RMP}, where the
solitons are always confined in pairs.  In cases of continuous symmetries, the
ordering violated by the amplitude kink can be restored by changes in the
phase of the complex order parameter $\Psi(x)$, localized in the tails of a soliton of
length $l_{phase}\sim T_{c}^{-1}$, where $T_c$ is the temperature of the long-range ordering due to the interchain coupling. But universally, except truly $1D$ systems
like isolated atomic chains \cite{Chao-NC:2016}, there is a local lifting of
degeneracy that comes from interchain interactions, which are
responsible for establishing the long-range $2D$ or $3D$ ordering. Namely, the interchain ordering
energy $J_{\perp}$ (per longitudinal lattice unit of the length $a_{\parallel}$) is paid when
adjacent domains at neighboring chains are not rightly correlated.

The confinement interaction determines the intermediate phases and the
kinetics of aggregation of nonequilibrium (e.g., optically induced) domains
into the long-range ordered phase.

\subsection{Qualitative description of phase transitions in the ensemble of neutral solitons}
\label{Tentative_phase_diagram}

\begin{figure}[tbh]
\centering
\subfloat[\label{a}]{%
  \includegraphics[width=.45\linewidth]{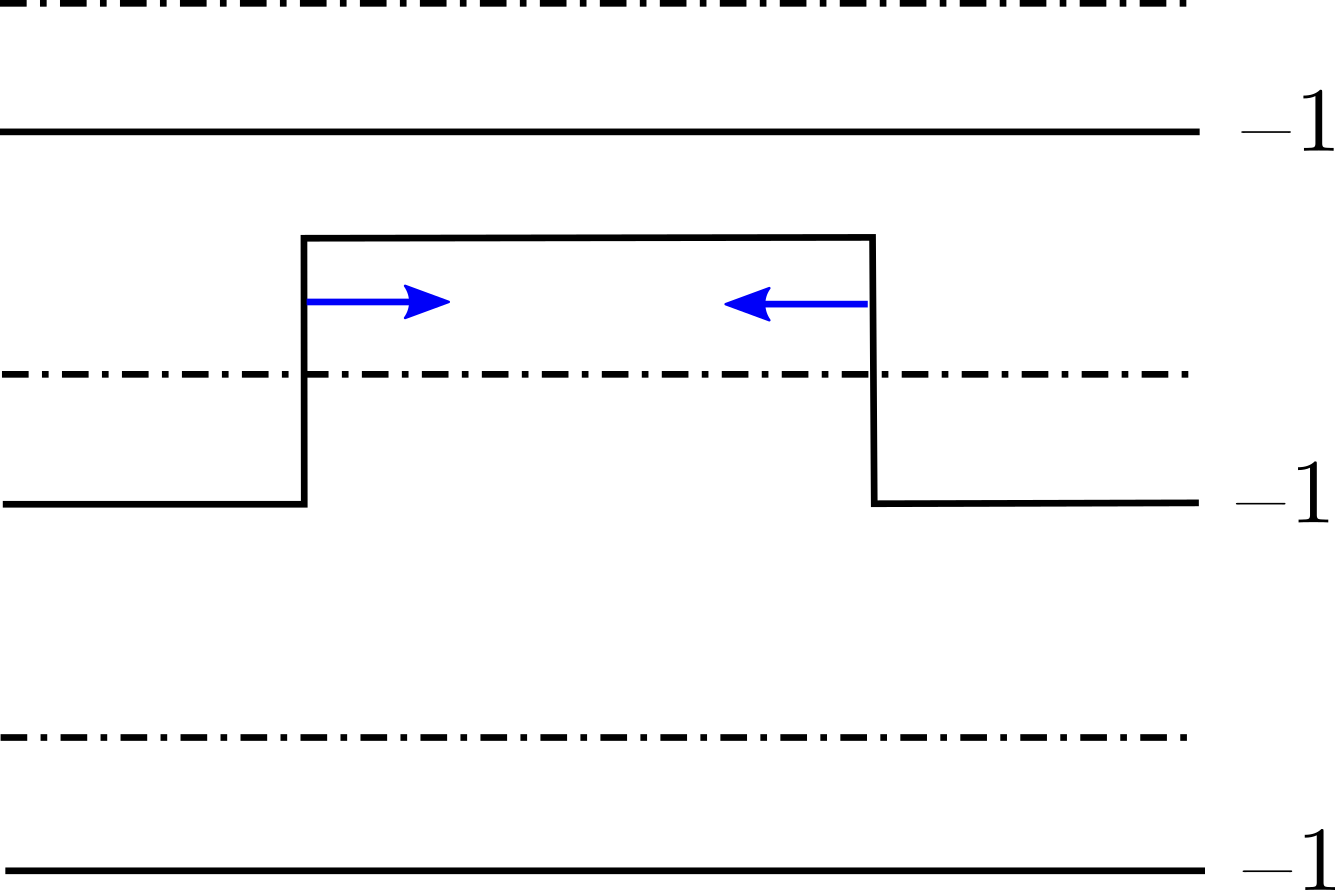}%
}
\hfill
\subfloat[\label{b}]{%
  \includegraphics[width=.45\linewidth]{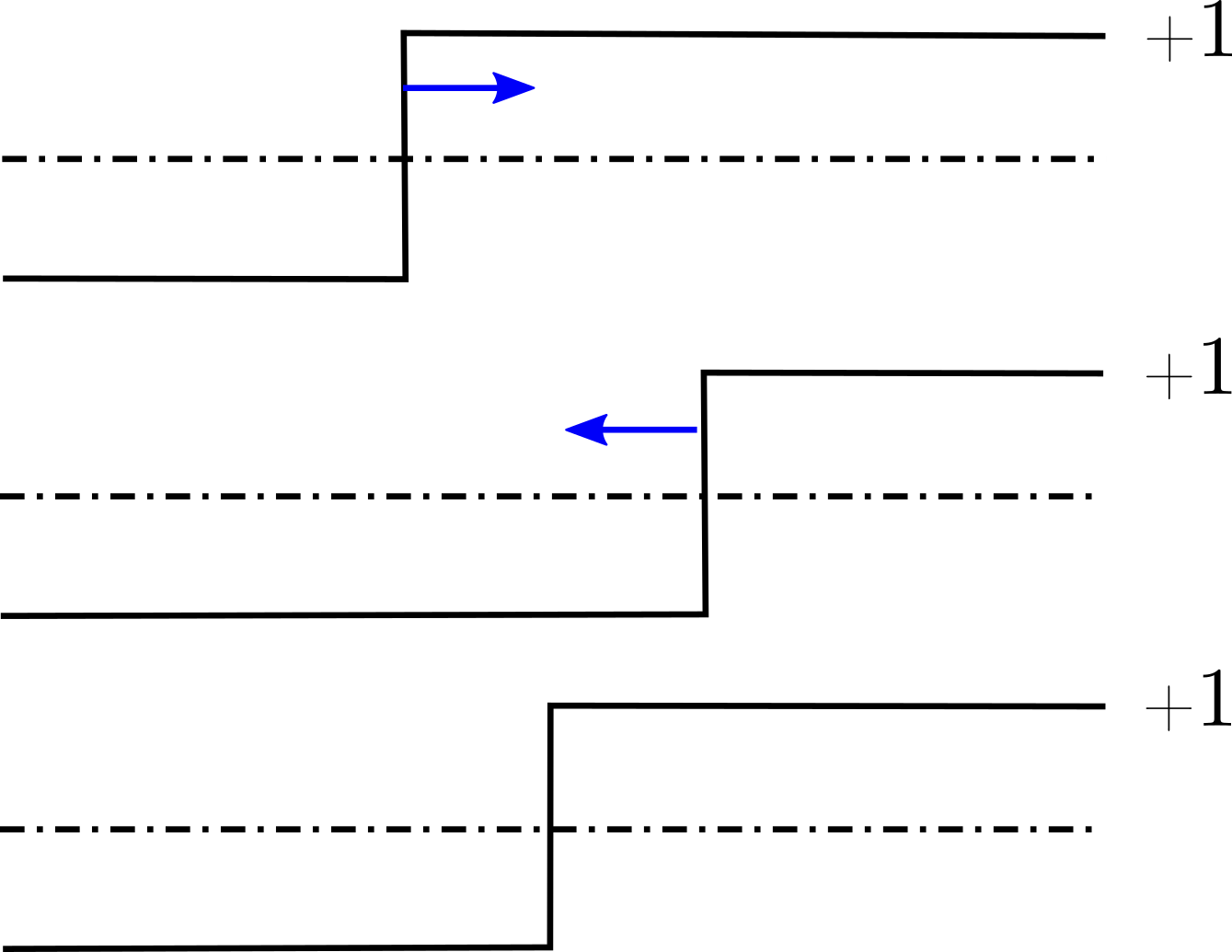}%
}
\caption{ (Color online) High temperature $T>T_{1}$: Solitons
exist as individual entities. They already experience long-range attraction
towards binding them into pairs at the same chain (a) or to walls at
neighboring chains (b). Horizontal black lines correspond to ground states with the order parameter $\pm 1$, vertical lines represent kinks,
arrows show forces acting upon the kinks.}
\label{fig:T>T1}
\end{figure}

Consider the influence of weak ordering interaction between chains ($2D$ or $3D$
coupling) on the state and statistical properties of the kinklike solitons
\cite{Bohr:1983}. The weak interchain coupling does not affect substantially
the structure of the soliton core, but below the $2D$ or $3D$  ordering temperature
$T_{1}$, its role turns out to be fundamental at large distances. Since
each kink separates different states of the system, the correlation between
the chains is violated in its vicinity. As a result, the system loses an
energy $2 J_{\perp}$ per site, which increases proportionally to the distance
 from the soliton. As we see from Fig.\ref{fig:T>T1},
the energy grows both with separation of two solitons on one chain and among
solitons at neighboring chains, hence the tendencies to either formation of
on-chain bikinks or the interchain aggregation of solitons into walls. As a result of
these contradictory tendencies, as temperature lowers, the system passes through two phase
transitions: the coupling of solitons into pairs at $T=T_{1}$, and aggregation
of pairs between the chains at lower $T=T_{2}\ll T_{1}$. For a
three-dimensional system, the temperature $T_{2}$ is a phase-transition point,
below which plane domain walls appear in the system passing through the
entire cross section. For a two-dimensional system, this is not the distinct
phase transition; instead a gradual increase of the transverse dimension of the
paired walls takes place at $T<T_{2}$. For a finite system, like in our modeling, there is a sample dependent temperature $T_F$ where the first
wall crosses the whole sample even in $D=2$.

\begin{figure}[tbh]
\centering
\subfloat[\label{a}]{%
  \includegraphics[width=.45\linewidth]{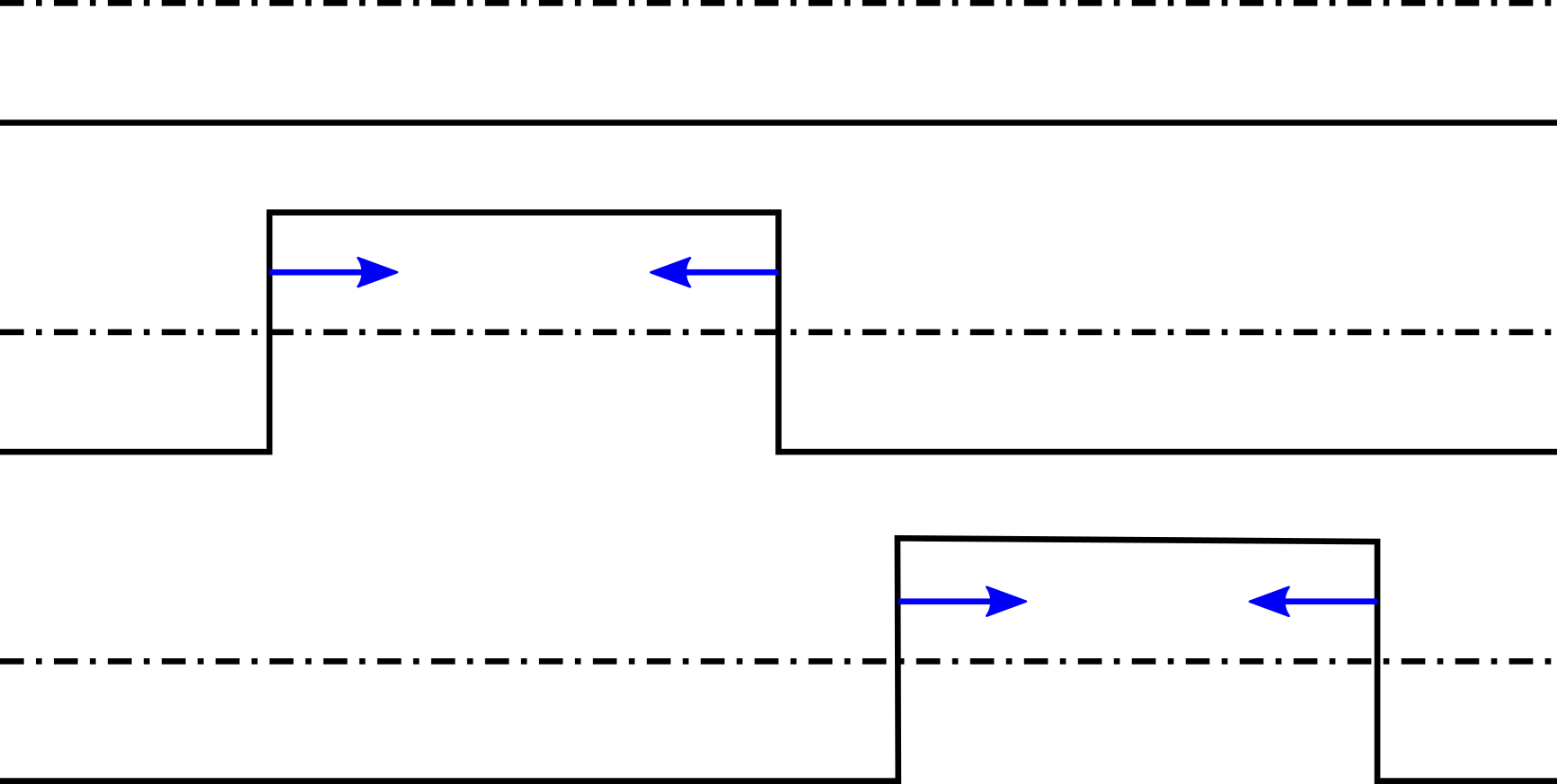}%
}
\hfill
\subfloat[\label{b}]{%
  \includegraphics[width=.45\linewidth]{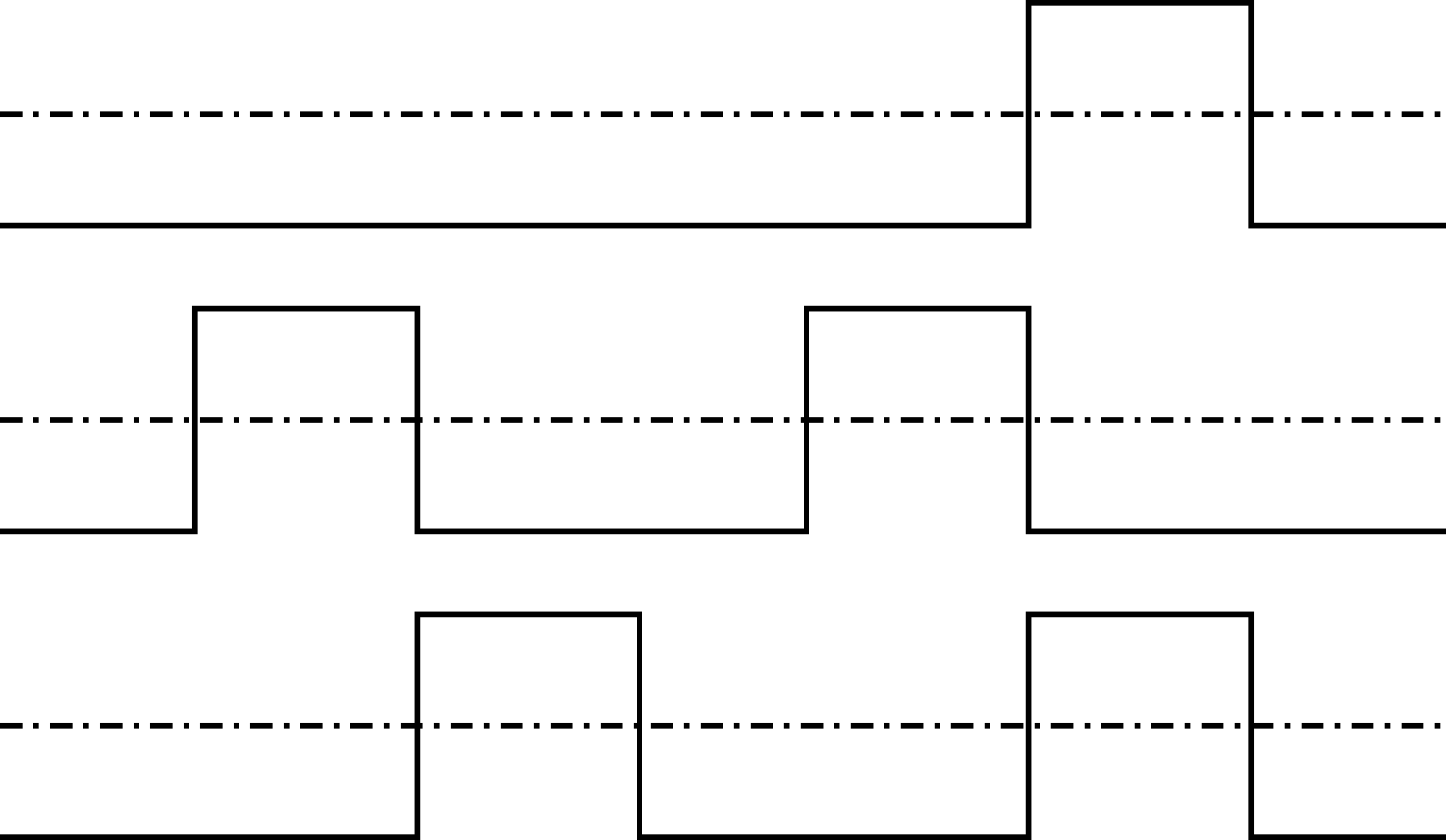}%
}
\caption{(Color online) Intermediate temperatures within
$T_1>T>T_2$. There are no more individual solitons, but a gas of their confined pairs. The
pair lengths are loose and fluctuate at $T_{1}>T>J_{\perp}$ (a); they are tight at
$J_{\perp}> T >T_2$ (b).}%
\label{fig:T1>T>T2}%
\end{figure}

\begin{figure}[tbh]
\centering
\subfloat[\label{a}]{%
  \includegraphics[width=.45\linewidth]{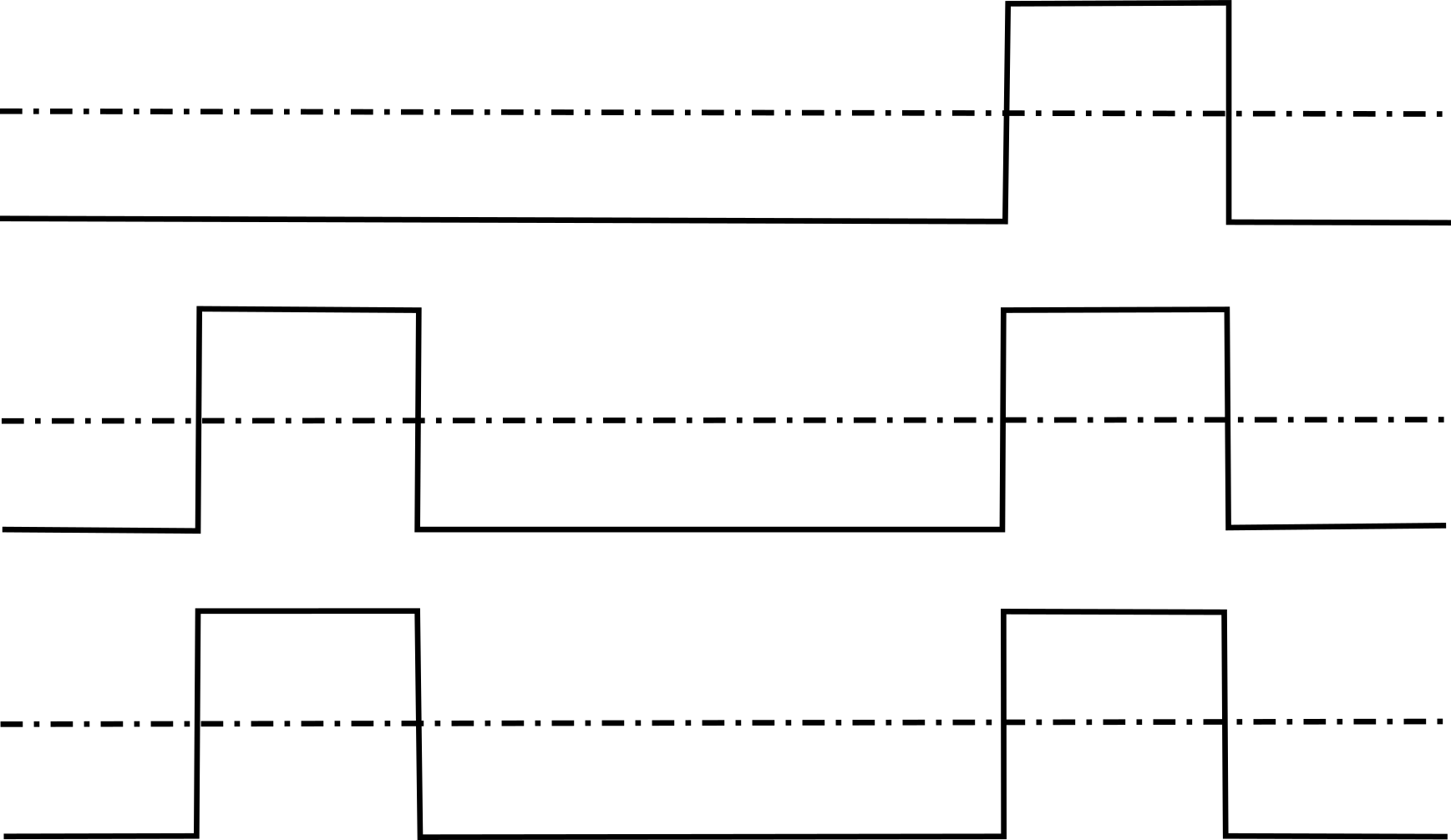}%
}
\hfill
\subfloat[\label{b}]{%
  \includegraphics[width=.45\linewidth]{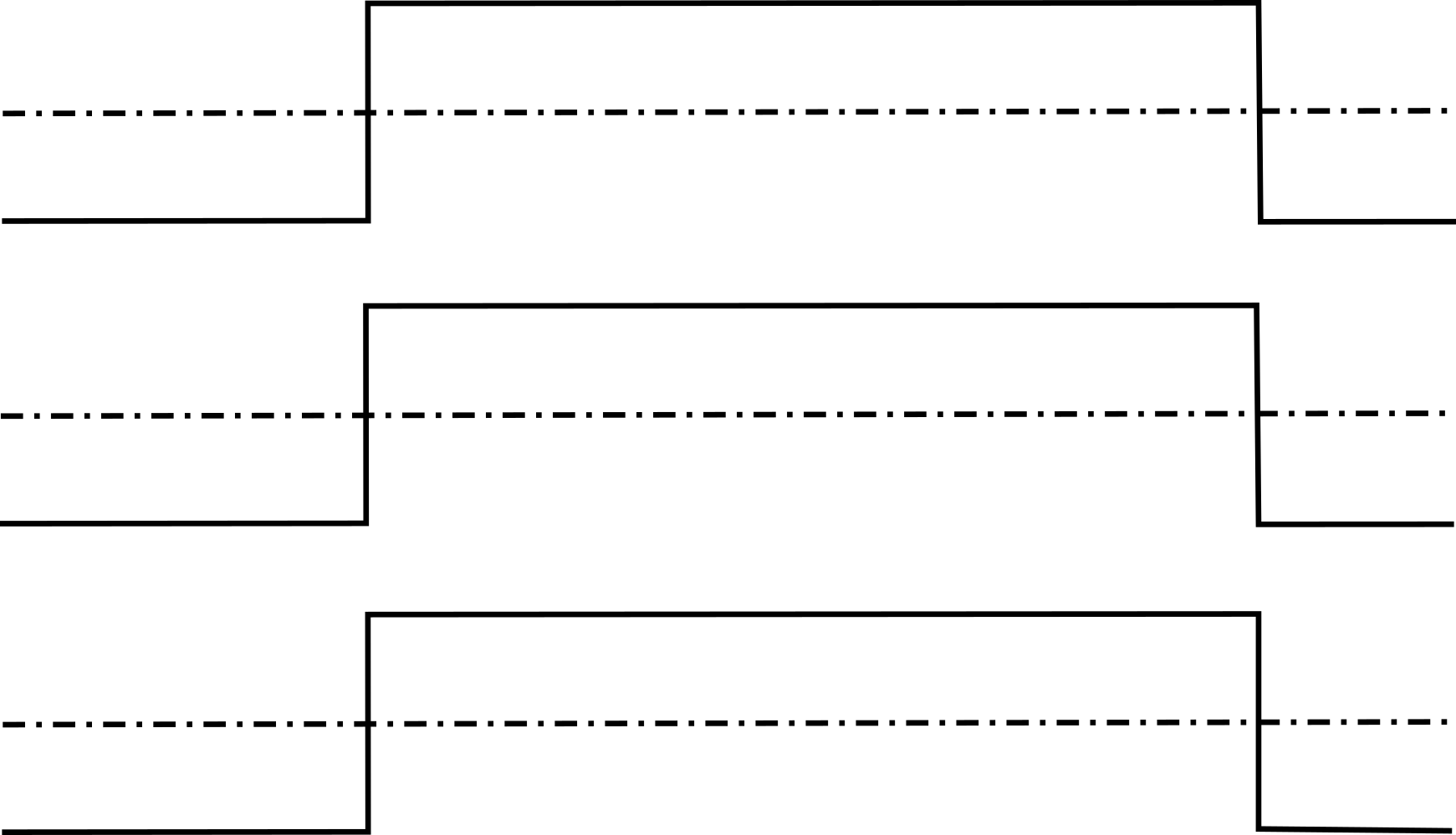}%
}
\caption{Low temperatures $T<T_{2}$: aggregation
of pairs into growing bikink walls (a) and then their disintegration into isolated walls of kinks (b).}
\label{fig:T<T2}%
\end{figure}

This qualitative picture is based upon an exact solution available
for a $2D$ system of neutral solitons with some qualitative extensions to the $3D$
case \cite{Bohr:1983}. The case of charged solitons has been addressed in \cite{Teber:2001, Teber:2002}, but with a restrictive constraint: the bisoliton pairs were not supposed to move from one chain to another.
Here we shall present the results of unrestricted numerical modeling performed for the challenging case of a
$3D$ system, both for neutral and charged ensembles of solitons.

In short, the picture of equilibrium states is the following. With lowering $T$
at a given concentration of kinks $\nu$, the system passes through a
sequence of two phase transitions (at $T=T_1,T_2$) and a crossover in between (at $T=J_{\perp}$), which are determined by three scales of
energy: $J_{\perp}$ as the local energy of the interchain ordering, the bigger $T_1 \propto J_{\perp}/\nu$
as the nonlocal energy of soliton confinement, and the smaller $T_2 \propto J_{\perp}/\ln(1/\nu)$ as the characteristic energy of soliton
transverse aggregation.

I. $T_{1}\propto Fl \propto J_{\perp}/\nu$ (see Fig.\ref{fig:T1>T>T2}). For the order parameter, this is the conventional $3D$ ordering transition realized in a quasi-$1D$ system.
However, for the ensemble of solitons, this is a confinement transition, which takes place when the temperature decreases below the mean
interchain interaction $Fl$ at the mean distance $l = a_{||}/\nu$ between the solitons.
Below that, at $T<T_{1}$, individual on-chain solitons cannot exist, they are
confined into loose pairs which form strings of size $l_{bs}$ with energy $F l_{bs}$ among them ($T_1$ can be also defined by the
condition $l_{bs} \sim l$, when confined pairs dissociate).
With lowering $T$, the pairs become progressively more confined at the thermal
length $l_{T}=T/F \ll l$, with rare collisions among the pairs.
At $T \ll J_{\perp}$, the pairs become tightly bound, their spacing
does not fluctuate anymore.
The confinement energy is reduced from the high-$T$ scale $J_{\perp}$ per unit cell to $J_{\perp}$
per kinks' pair.
Actually, the pair length $l_{bs}$ shrinks from the
thermal one $l_{T}$ to the quantum zero point limit $l_{q}$ such that $Fl_{q}%
\sim\hbar^{2}/Ml_{q}^{2}$ ($M$ is an effective mass of the soliton).

Thus, the energy $T_{1}$ has two faces. For the order parameter, this is just the
transition temperature $T_{1}=T_{c}$ of the second-order phase transition
to the state with its nonzero mean value at
$T<T_{c}$. However, for solitons, this is the confinement transition temperature, below which
they become bound into pairs -- the bisolitons.

II. $T_{2}\propto J_{\perp}/\ln(1/\nu)<J_{\perp}$ (see Fig.\ref{fig:T<T2}). This temperature
can be viewed as the onset of aggregation of solitons when first domain walls appear
crossing the whole sample or forming macroscopic bubbles.  As the temperature lowers beyond $J_{\perp}$, the bisolitons start to aggregate in
transverse disks, and finally at $T_2$ these disks cross the entire sample. Aggregation to
domain walls gains the confinement energy, which now is not lost at all -- the
neighboring chains are always in the right arrangement. However, the entropy is
lost, and this balance determines $T_{2}$. The pairs still coexist with walls
below $T_{2}$, but with further decreasing $T$ they vanish providing the material for
building more macroscopic walls.

The $T_{2}$ transition can be viewed similarly to a vapor condensation in a
given volume when the first appearing wetting fixes the chemical potential
(the saturation pressure) of the gas. Another analogy is the Bose-Einstein condensation, but in real, instead of the reciprocal, space. Indeed,
below $J_{\perp}$, there is the gas of "confined pairs"  (bisolitons) with energy $W_{bs}=F l_{bs}$, whose chemical potential $\mu_{bs}$ is adjusted to maintain the given total concentration
$\nu_{bs} (\mu_{bs},T)=\exp((\mu_{bs}-W_{bs})/T)=\nu/2$. When the first macroscopic domain walls appear,
they serve as a reservoir of kinks fixing their chemical potential -- ideally
at $\mu_{s}=0$, hence $\mu_{bs}=2\mu_{s}=0$. Then the concentration of pairs is
$\nu_{bs}(0,T)=\exp(-W_{bs}/T)$, and this number falls below the total available
$\nu_{bs}=\nu/2$, which happens at $T<T_{2}$.
The deficit $\delta \nu=\nu-2\nu_{bs}(0,T)$ gives the number of solitons that have been used to build the domain walls, with $a_{||}/\delta \nu$
giving the mean period of the stripe array
of domain walls. Notice a curious behavior: with lowering $T$ below $T_{2}$,
the mean value of the order parameter over the bulk disappears, while it is
present over each cross-section. The $T_{2}$ transition is the one
where the effective dimensionality of the system $D$ is reduced by $1$! We shall
see this explicitly via the effective Ising model with constraints.

\subsection{Evolution after the optical pumping.}

\begin{figure}[tbh]
\includegraphics[width=1.0\linewidth]{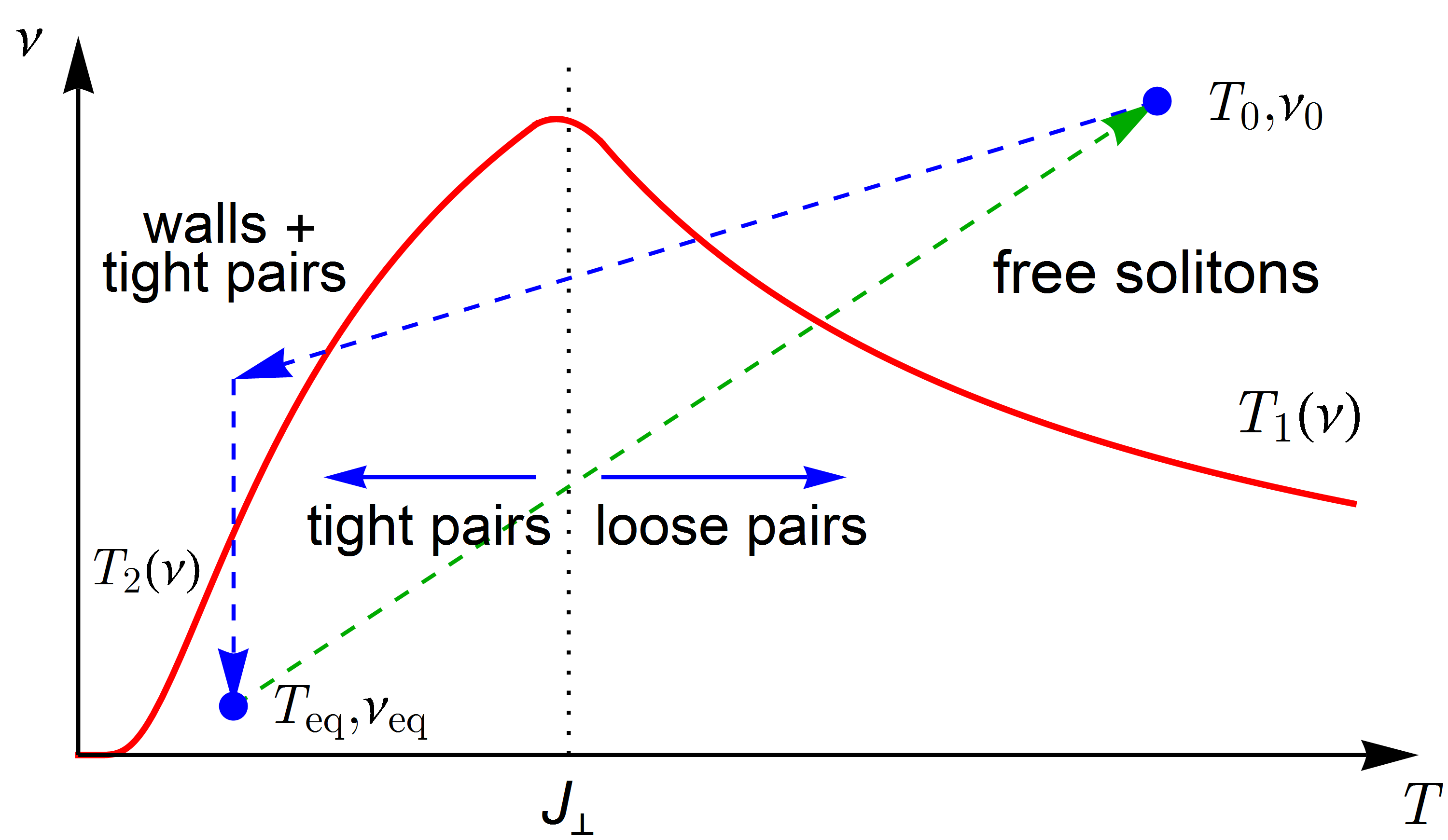}
\caption{(Color online)  The phase diagram of the soliton
ensemble in variables temperature vs concentration. Thick solid lines show the phase
transitions $T_{1}$ and $T_{2}$, the vertical dashed line shows a crossover at
$T\approx J_{\perp}$, the dashed arrowed lines give schematically the trajectory of
the pumping and subsequent relaxation (decrease of the total $\nu$ and the cooling).}%
\label{fig:phase-diag}
\end{figure}

After the optical pulse has created the ensemble of solitons, their mean
concentration $\nu$ starts to evolve as $\nu(t)$ both to thermal
equilibrium at a given $T$ and also together with the temperature $T(t)$. On
this way, the system will pass through a sequence of phase transitions or at
least of crossovers among different regimes which have been classified above.
The trajectory is summarized in Fig.\ref{fig:phase-diag}.

Before the pumping, at the equilibrium ambient temperature $T_{eq}$, the solitons are
present either as free particles, if $T_{eq}>T_{1eq}$, or as confined pairs if
$T_{eq}<T_{1 eq}$, where $T_{1 eq}=T_1(\nu_{eq})\approx
E_s/\ln(E_s/ J_{\perp})$ is the ordering transition temperature without
pumping (here $E_s$ is the soliton's core energy). However, in any case, the equilibrium concentration of kinks is low: $\nu_{eq} \sim
\exp(-E_{s}/T)$ at $T_{eq}>T_{1eq}$ or $\nu_{bs}\sim\exp(-2E_{s}/T)$ at
$T_{eq}<T_{1eq}$, because now the total big activation energy $E_{s}$ of the
soliton is to be paid in comparison with the relatively small scale of the
interchain energy $J_{\perp}$ when the number of solitons is controlled.

Just after the pumping, the initial concentration of kinks $\nu_{0}$ is very
high and the initial temperature $T_{0}>T_{eq}$ is also high (it can even
further increase at intermediate times because of the energy release from
relaxation of excitations \cite{Okamoto-NI}). Suppose that
$\nu_{0}>J_{\perp}/T_{0}$ so that we are in the disordered phase above $T_1(\nu_0)\propto  J_\perp/\nu_0$ -- the
pumping has destroyed the long-range order, the solitons are not confined in
pairs. With time, both $\nu(t)$ and $T(t)$ decrease, so the two sides of the
last inequality move towards each other, the transition is inevitably reached
at some time $t_{1}$ when $\nu(t_{1})=J_{\perp}/T(t_{1})$. Recall that this
temperature is well below the thermodynamical transition temperature $T_{1 eq}$ since the
number of solitons is still strongly enhanced.

The situation looks, at first sight, a bit less certain for the lower
transition because its expected temperature $T_{2}(\nu)=J_{\perp}/\ln(1/\nu)$
falls with decreasing $\nu(t)$, unlike $T_1(\nu)$. However, $T_{2}(\nu)$ decreases very slowly,
e.g., $\sim1/t$ for the exponential decay of $\nu$, which is slower than any
expected decrease of $T(t)$. So the second phase transition will also happen
at a certain moment $t_{2}$ of time, when $T(t_{2})=J_{\perp}/\ln(1/\nu(t_{2}))$,
then the confined pairs of solitons will start to aggregate into macroscopic
domain walls of solitons. The $T_2$ transition line will be crossed back (via evaporation of domain walls) in the course of returning the
temperature to the ambient one and the full annihilation of excess solitons.

    \section{The basic model.}
	
\subsection{Mapping to the constrained Ising model.}
	
For a generic (no CIs) model, the configurational energy for the order
parameter $\eta_{\alpha}(x)$ at the point $x$ of the chain $\alpha$ is
\begin{align}
H_0 =  \int dx\left( - \sum_{\langle \alpha ,\beta \rangle} V_{\perp}
\eta_{\alpha}(x)\eta_{\beta}(x) +  \right. \nonumber \\
+ \sum_{\alpha} \left( U(\eta_{\alpha}(x)) + C (\eta'_{\alpha}(x))^2 \right) \left. \vphantom{\sum_{\alpha}} \right)
\label{H0}
\end{align}
\noindent where $V_{\perp}$ is the interchain ordering energy per unit longitudinal length (actually, $2 Z V_{\perp}$ is the
confinement force $F$, where $Z=2,4$ is the number of nearest neighboring chains for  $D=2,3$) and $U(\eta)$ is a double-well potential with two
symmetrical minima normalized to $\pm 1$, which determines the two possible
equivalent ground states. The soliton is a trajectory, taking the
energy cost $E_{s}$, which commutes between these two minima; e.g. $\eta (x)=\pm \tanh
(x/a)$ or whatever is the antisymmetric solution determined by the competition of second and third terms in (\ref{H0}).
It is convenient to quantize the chain length as $x\Rightarrow x_{n}=na_{||}$ in some units $a_{||}$
well exceeding the intrinsic width $a$ of the soliton (we take $a_{||}$  to be equal to the quantum zero point limit size $l_q$ -- the minimal size of a bisoliton, see Sec. \ref{Tentative_phase_diagram}) and to introduce
the Ising spin variable as $S_{n,\alpha }=\eta _{\alpha }(x_{n})$. Then the
solitons, with the linear density $c_{\alpha}(x)$, are seen as sharp kinks,
whose number per site is given by the lattice function $\rho_{n,\alpha }$ such that \cite{Bohr:1983}
\begin{equation}
\begin{split}
c_{\alpha }(x)a_{||}&=\rho _{n,\alpha }
=\frac{1}{2}(1-S_{n,\alpha }S_{n+1,\alpha }), \\
\left\langle \rho_{n,\alpha}\right\rangle &= \nu = N_{s} /L H^{D-1},
 \label{rho}
\end{split}
\end{equation}
where $\nu $ is the mean concentration of solitons per site, $N_{s}$ is their
total number, $L\times H^{D-1}$ are the dimensions of the sample in units of $a_{||}$ and $a_{\perp}$, respectively.
Representation (\ref{rho}) underlines the fact that a soliton is present at the site $n,\alpha$ of the dual lattice only if $S_{n,\alpha} = 1$ and $S_{n+1,\alpha} = -1$ or vice versa.
In the single chain limit, $S_{n}$ and $\rho _{n}$ play the roles of complementary order and disorder parameters (see, e.g., \cite{Tsvelik:2003}). The mean density can be controlled by the chemical potential
$\mu_{s}$ and we arrive at the Gibbs energy $\widetilde{H}_{0}$ for the grand
canonical ensemble of solitons \cite{Bohr:1983}:
\begin{equation}
\begin{split}
\widetilde{H}_{0} &= H_{0}-\mu_{s} N_{s} =  \\
& = - V_{\perp}\sum_{\langle \alpha ,\beta \rangle} \int dx \, \eta_{\alpha}(x)\eta_{\beta}(x) +(E_{s}-\mu_{s})N_{s} = \\
&=-J_{\perp }\sum_{\langle\alpha ,\beta \rangle n}S_{n,\alpha }S_{n,\beta }
-J_{||}\sum_{\alpha,n}(S_{n,\alpha }S_{n+1,\alpha }-1);   \\
\text{where} &\,\,\, J_{\perp} = V_{\perp}a_{||} ~,~J_{||}=(E_{s}-\mu_{s})/2.
\label{originalH}
\end{split}
\end{equation}
In a number of cases, particularly in application to doping, the solitons can possess an electric charge. If screening by external carriers
is strong enough, these solitons behave as neutral ones. However, in case of intermediate or weak screening, the charges of solitons must be
considered explicitly.
Therefore, taking into account the CI, we can add also \cite{Teber:2001} the Coulomb energy $H_{C}$ to the Hamiltonian:
\begin{equation}
\widetilde{H}=\widetilde{H}_{0}+H_{C}~,~H_{C}=\frac{e^{2}}{2\epsilon}
\sum_{n,m;\alpha ,\beta }\frac{(\rho _{n,\alpha }-\nu )(\rho _{m,\beta }-\nu
)}{|{\mathbf{r}}_{n,\alpha }-{\mathbf{r}}_{m,\beta }|},  \label{HCoulomb}
\end{equation}
where $e$ is the electron charge and $\epsilon$ is the dielectric constant,
taken here to be isotropic. It is known that competing short-range attractive and long-range repulsive forces may result in the formation of a
diverse variety of patterns \cite{Muratov:2002,Kabanov:2005,Kabanov:2008,Zhang:2011}. In Sec. \ref{Numerical_approach}, we shall describe
Monte Carlo modeling for both neutral and charged cases.

It is remarkable that controlling the chemical potential, the grand
canonical ensemble of solitons in a $D$-dimensional system can be described by
the $D$-dimensional Ising model, while the canonical ensemble is described by the stack of noninteracting $(D-1)$-dimensional models
with an overall constraint. In this anisotropic model, only the interchain coupling constant $J_{\perp}$ has a physical origin and is frozen,
while
the on-chain constant $J_{||}$ is determined by the chemical
potential.  Since the physical situations of interest correspond to
controlling the concentration $\nu$, then the price, and the source, of the most
interesting behavior come from the self-consistency condition to invert
the function $\nu (\mu ,T)$ to $\mu (\nu ,T)$, hence traveling over a special
line on the surface of Ising model parameters. In this way, we can take a
good advantage (without CI) of notion on the Ising models (full
for $D=2$ and qualitative  at least for $D=3$). On the other hand, as we shall
demonstrate in the following, a numerical procedure can be constructed
which allows to directly keep the constraint on the total number of
solitons. Then working with the canonical ensemble we get a good advantage
to deal with  $(D-1)$-dimensional systems with only physical interchain interaction being
present.

\subsection{Estimations based on the effective Ising model for a neutral system}
\label{Estimations_based_on_the_effective_Ising_model_for_a_neutral_system}

Consider the effective Ising model with adjustable $J_{||}=J_{||}(T,\nu )$
in $D$ dimensions. The concentration of solitons is always supposed to be
small $\nu \ll 1$.
Here we find the adjusted values of $J_{||}(T,\nu)$
in $2D$ and $3D$ for limiting cases of high and low temperatures and make estimations for $T_1(\nu)$ and $T_2(\nu)$.

In the high-temperature limit $T\gg T_{1}\gg J_{\perp }$,  the system is effectively one-dimensional;
the energy of one soliton is $2J_{||}$ and the probability to find a soliton at
a given point of the dual lattice is $\exp (-2J_{||}/T) \approx \nu,$ then
\begin{equation}
J_{||}(T,\nu)\approx \frac{T}{2}\ln \frac{1}{\nu}
\end{equation}
is the effective on-chain coupling,
hence, from (\ref{originalH}), the soliton chemical potential increases with decreasing $T$. Extrapolating this expression down to
$T_1$, we find an estimate for $T_1$.
For $D=2$, we deduce from Onsager's exact result \cite{Onsager:1944} $\sinh (2J_{\perp}/T_{1})\sinh (2J_{||}/T_{1})=1$  that $T_1 \sim
2J_{\perp}/\nu$, which up to a numerical factor, agrees with the result of Ref. \onlinecite{Bohr:1983}: $T_1 \approx 2J_\perp/\pi\nu$ (see also Appendix A).
For $D=3$, we can use the approximation, where the on-chain interaction is
treated exactly while the interchain one is taken into account using the
mean-field theory. For the critical temperature of anisotropic $3D$ Ising model, this approach gives \cite{Ising3DAnisotr1,Ising3DAnisotr2}
$T_{1}\approx 8J_{\perp }\exp \left( 2J_{||}/T_{1} \right)$, from which we get $T_1 \sim 8J_\perp/\nu$.
We see that both in $2D$ and $3D$ the
Ising critical temperature behaves as $T_{1}\sim J_{\perp}/\nu $ at $\nu \rightarrow 0$, which justifies treating $T_{1}$ as the confinement
transition.

Consider now the opposite limit of low temperatures $T\ll J_{\perp}$. It looks, at first sight and wrongly, that at low temperatures, the system
persists in a very simple form of one spin-ordered domain impregnated by a
dilute gas of spin-reversed sites, which are our tightly bound bisolitons with
the energy $2ZJ_{\perp}$. Taking into account the chemical potential and neglecting the excluded volume corrections, the concentration of bisolitons is
\begin{equation}
\nu_{bs}=\exp \left( -\frac{4 J_{||}+2ZJ_{\perp }}{T} \right)
\end{equation}
Since this number is fixed at $\nu _{bs}=\nu/2$, then
lowering $T$ must be compensated by the decrease of $J_{||}$,
which is limited to be positive $J_{||}\geq 0$: a negative $J_{||}<0$ would switch the system to an "antiferromagnetic"
ground state with spins alternation at each site in the chain direction,
hence the infinite number of kinks. Hence a new reservoir for the storage of
solitons must be opened when $T$ falls below $T_{2}(\nu)$ such that $\nu
=2\exp (-2ZJ_{\perp }/T_{2})$, i.e.,
\begin{equation}
T_{2} \approx \frac{2ZJ_{\perp }}{\ln (2/\nu)}  \label{T2-3D}
\end{equation}%
which agrees with both the estimation (\ref{AppA_T2}) for $2D$ (with logarithmic accuracy) and with the exact solution (\ref{T2}) accessible in
$3D$.
This new reservoir can be viewed as a system of stripes (lines in $2D$ or
planes in $3D$) which cross the whole sample separating the bulk in
noninteracting domains of alternating magnetization. A more thorough
analysis \cite{Bohr:1983} shows that $T_{2}$ must be indeed a sharp phase transition at $D=3$,
while at $D=2$ it is only a crossover for growing rods of a finite extent.
Only for a finite $2D$ system of width $H$ the rods pass through the entire
 cross-section of the sample at $T_{F}\approx 4J_{\perp }/\ln (H/2\nu)$.

Estimates for $J_{||}(T,\nu)$ can also be found at $T \ll J_{\perp}$. For $D=2$ as $T\rightarrow 0$ and $D=3$ as $T\rightarrow T_2+0$ we find
that $J_{||} \rightarrow 0$ as (details of derivations are given in the Appendix A)
\begin{align}
\qquad J_{||}(T,\nu) & \propto  \, T \exp(-2J_{\perp}/T)/\sqrt{\nu}   &(2D), \label{JparLowT} \\
\qquad J_{||}(T,\nu) & \propto  \, \nu \ln(1/\nu) \cdot (T-T_2)       &(3D). \nonumber
\end{align}
Below $T_2$ in $3D$, transverse layers do not interact and $J_{||}$ remains $0$.

This picture, and its strong complication by long-range CIs will be numerically verified
and expanded in the next section. Some analytical results for both neutral and charged systems are also given in the appendices.

%__________________________________________________________________
%                      MONTE_CARLO SIMULATION
\section{Numerical approach}
\label{Numerical_approach}

\subsection{Monte Carlo simulation details}

In this section, we consider the Monte Carlo (MC) simulation of the ensemble of solitons. We have studied the statistical properties
of the canonical ensemble of solitons over broad temperature ranges in two and three dimensions for both neutral and charged solitons.

In the numerical simulations, it is more efficient to keep a fixed number of solitons rather than using self-consistent $J_{||}(T, \nu)$ as we did
in the previous section. Here then we shall work with the canonical ensemble of solitons keeping fixed their overall number, while the
number of solitons at a given chain can vary (in contrary to the more restrictive prescription employed in \cite{Teber:2001,Teber:2002}).
%Fluctuation of solitons' number $N_s$ for grand canonical ensemble is proportional to $\sim \sqrt{N_{s}}$ which means that in the limit of large $N_{s}$ these two approaches should give convergent results.

\begin{figure}[tbh]
\centering
\subfloat[Soliton moves along the chain  \label{a}]{%
  \includegraphics[,width=0.75\linewidth]{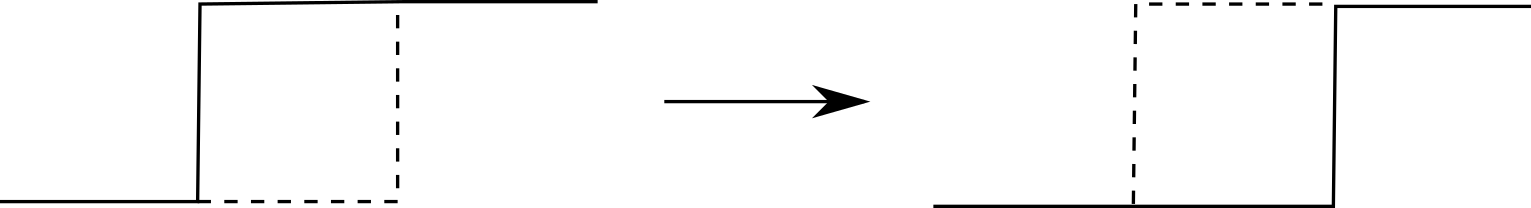}%
}

\subfloat[Bisoliton moves along the chain \label{b}]{%
  \includegraphics[width=1.0\linewidth]{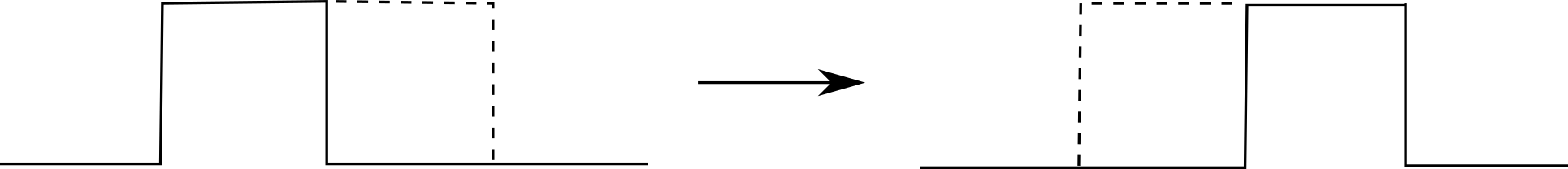}%
}

\subfloat[Bisoliton moves across the chain \label{c}]{%
  \includegraphics[width=0.75\linewidth]{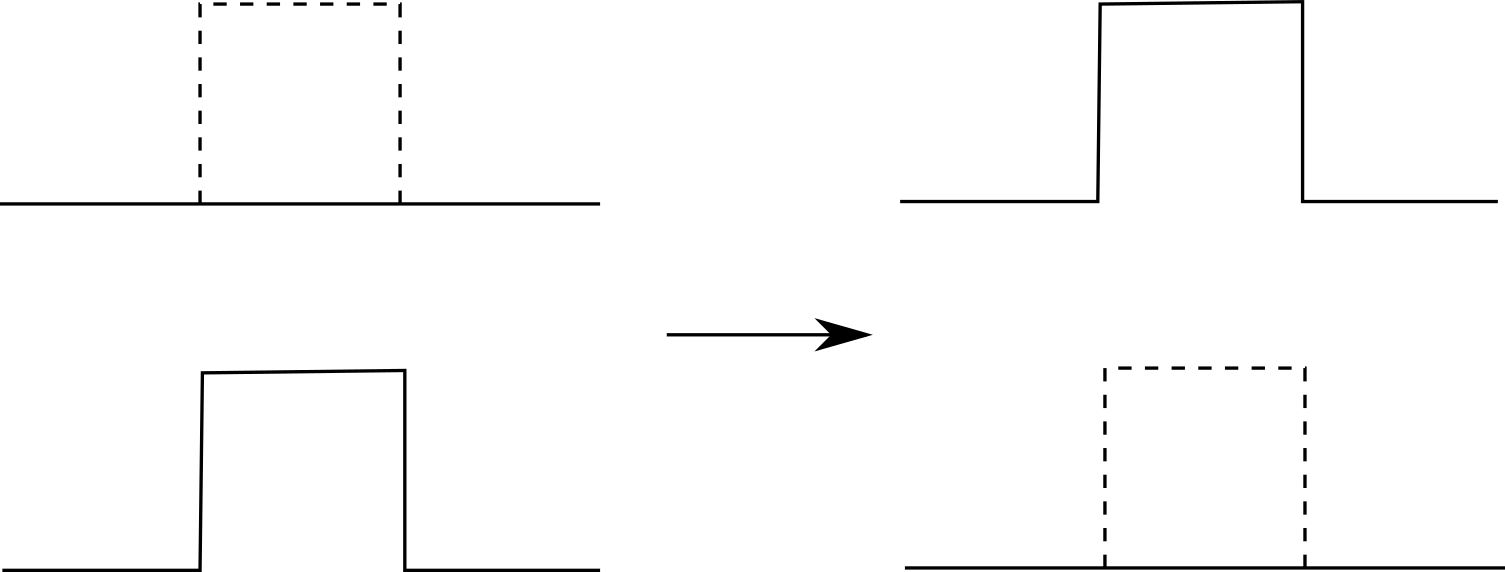}%
}
\caption{Three types of elementary MC movements (a-c).
Left panes show the state of the system before the movement and right panes -- after it. Dashed lines show the difference between old and new configurations.}
\label{SolMovements}
\end{figure}

The MC simulation was performed with the use of the standard Metropolis algorithm with three types of unit movements: moving a single soliton along
the chain [Fig.\ref{SolMovements}a], moving a solitons' pair along [Fig.\ref{SolMovements}b], or across the chains [Fig.\ref{SolMovements}c]. The
type-(b) movement is a superposition of two type-(a) movements, but it is useful to consider the type-(b) movement explicitly since it improves
the low-temperature acceptance rate of the algorithm.

For the case of charged solitons, we need to take into account the long-range CI, which is always a formidable task, and even more here for the
system prone to pattern formation, owing to the appearance of locally noncompensated charges at a growing scale.
Since we are interested in the wall formation process, and for an infinite charged wall the electric potential grows linearly
with the distance, then correctly imposing periodic boundary conditions for the CI becomes
very important.
We use the periodic version of the Coulomb Hamiltonian (\ref{HCoulomb}) (for simplicity, we put $a_{||} = a_{\perp}$ in the simulation),
where the summation goes over not only all pairwise interactions within a computational
cell, but also between solitons and their images and between images and the neutralizing
negative background. In practice, this is done by the technique of Lekner \cite{Lekner91} for both $2D$ [\onlinecite{Gronbech97}] and $3D$ [\onlinecite{Lekner98}] cases. This technique allows to efficiently
calculate the force acting on a given particle from another particle and all its images and, and by integrating the force, to obtain the effective
pairwise potential. Finally, we tabulate this potential for fast computations.

Since the CI is the long-range one, the calculation of the energy change $\Delta E_{C}$ at each MC-step can be time consuming.
To deal with this we use two different approaches. When the MC acceptance rate is high, we employ the first (standard) approach: at each MC trial
step we recalculate the CI energy of the shifted soliton with respect to other solitons. However, at low temperatures, when the MC acceptance
rate becomes low, it turns out that this approach is ineffective, since many calculations are wasted to compute $\Delta E_{C}$ for rejected
steps.
Therefore, at low temperatures (and low MC acceptance rates) we use another algorithm \cite{GrestXY,LeeXY}:
instead of recalculating the interaction energy of the shifted soliton with every soliton in the system at each trial MC step
(computational cost of which is $O(N_{s} \times N_{trial steps}$)), we introduce an electric potential $\phi$ and use it to calculate the
Coulomb energy change: $\Delta E_{C} \simeq e (\phi(r_{new})-\phi(r_{old}))$, the computational cost of which is only $O(N_{trial steps})$. However,
now we have to update the potential after every accepted step at each site of the system, which costs $O(Volume\times N_{accepted steps})$.
This means that this approach works better when the acceptance rate is low.
Combining these two approaches allows us to effectively perform simulations for the system with long-range CI at both high and low temperatures
and even in $3D$ space.

\subsection{Numerical results for 3D case}
\label{Numerical_results_for_3D_case}
In this section we shall describe our main results: the evolution of the system with lowering temperature in different regimes. We shall use several
presentations: images for distributions of solitons and spins. These pictures will be further characterized by plotting the numbers of
nearest spins or solitons.

\subsubsection{Condensation of solitons into walls for neutral solitons}
\label{Condensation_of_solitons_into_walls_for_neutral_solitons}

\begin{figure}[tbh]
\centering
\subfloat[Solitons, $T=2.1 J_{\perp}$ \label{2}]{%
  \includegraphics[width=0.94\linewidth]{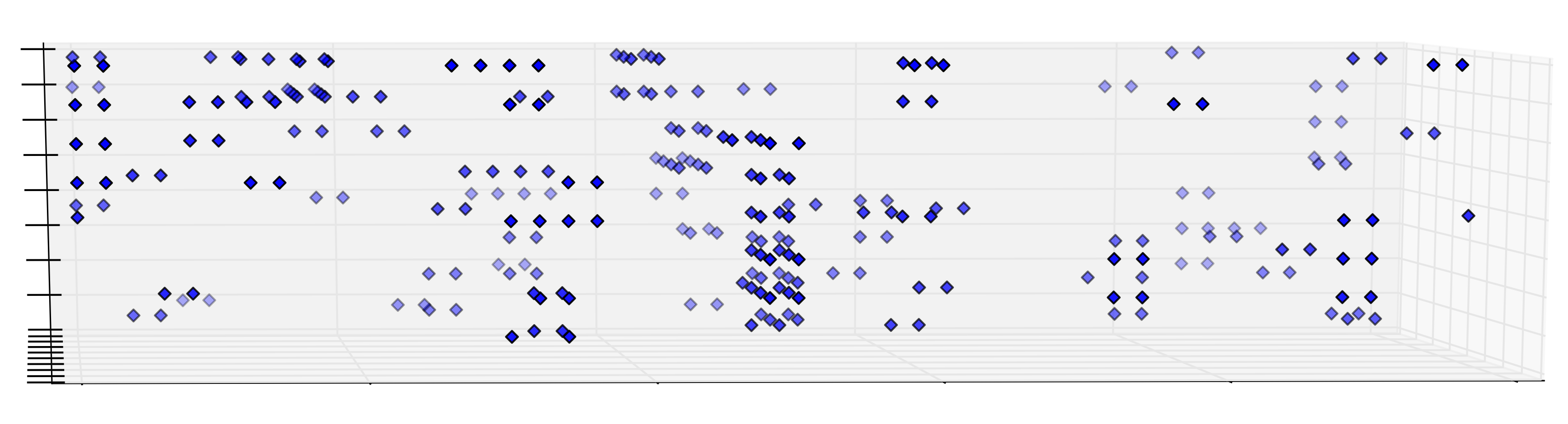}%
}

\subfloat[Ising spins, $T=2.1 J_{\perp}$ \label{1}]{%
  \includegraphics[width=0.94\linewidth]{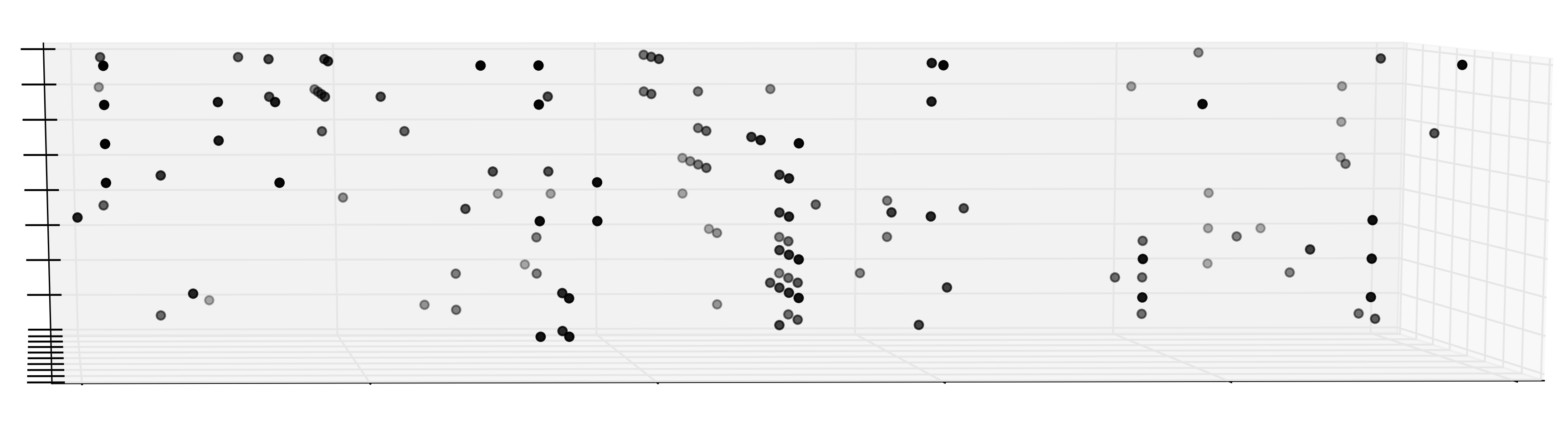}%
}

\subfloat[Solitons, $T=2.0 J_{\perp}$ \label{2}]{%
  \includegraphics[width=0.94\linewidth]{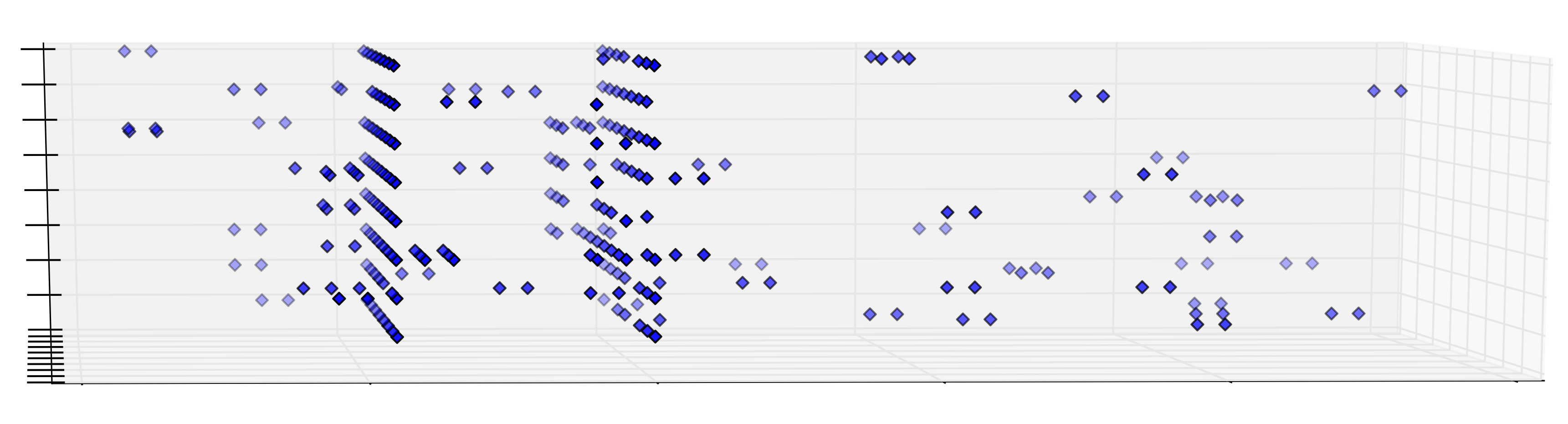}%
}

\subfloat[Ising spins, $T=2.0 J_{\perp}$ \label{2}]{%
  \includegraphics[width=0.94\linewidth]{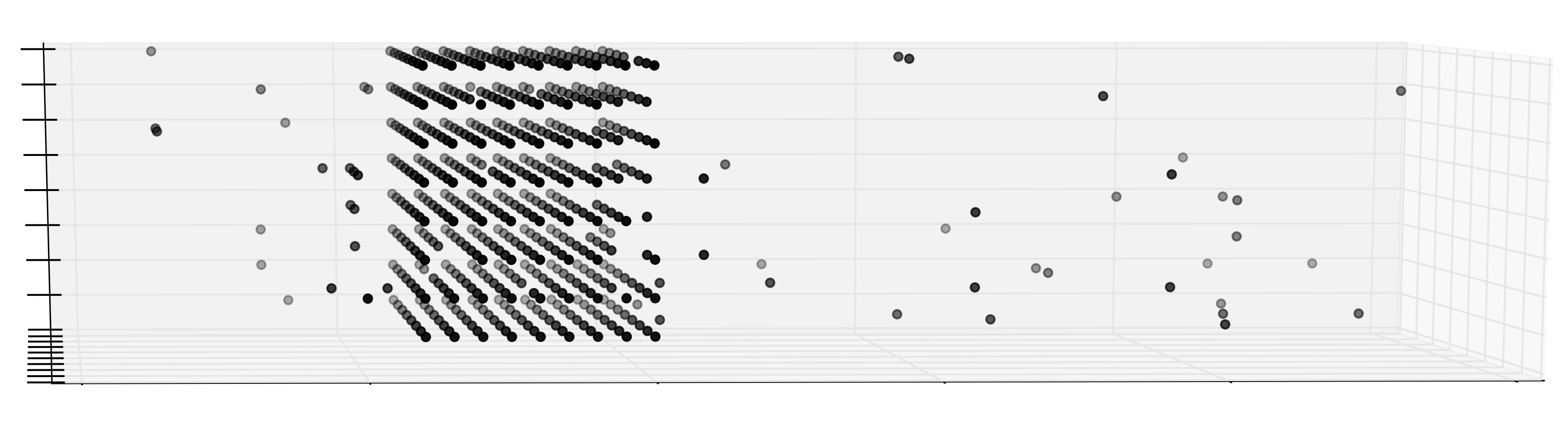}%
}

\subfloat[Solitons, $T=1.6 J_{\perp}$ \label{2}]{%
  \includegraphics[width=0.94\linewidth]{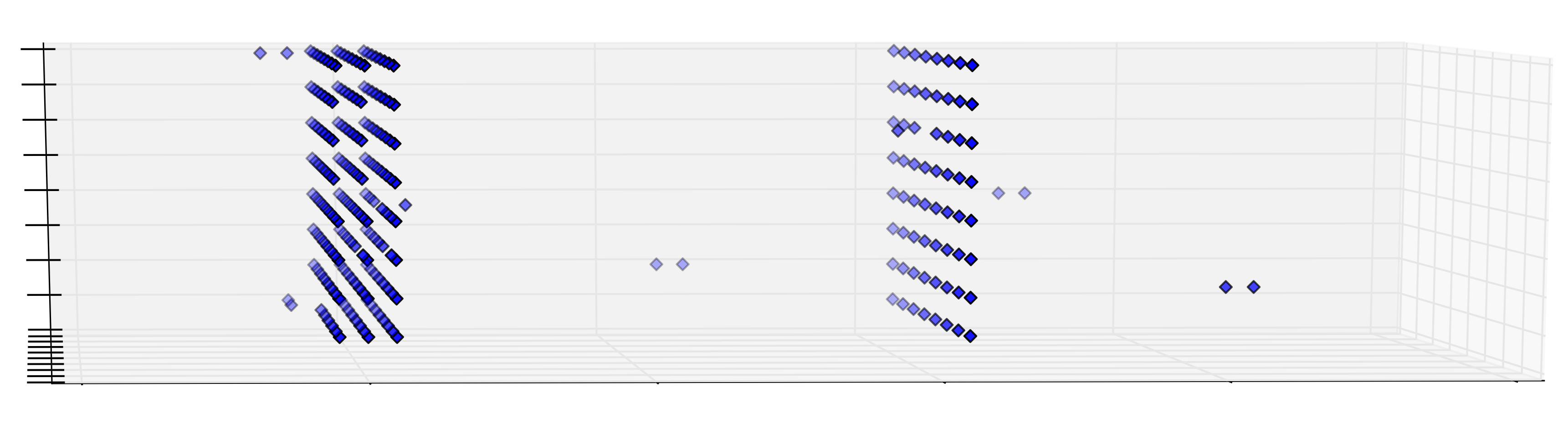}%
}

\subfloat[Ising spins, $T=1.6 J_{\perp}$ \label{3}]{%
  \includegraphics[width=0.94\linewidth]{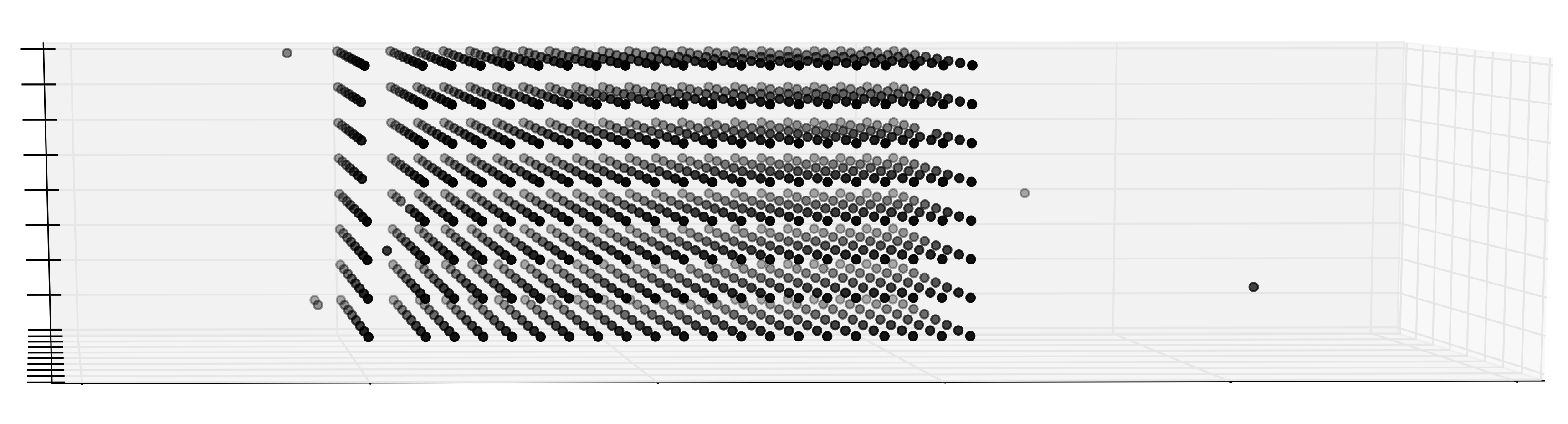}%
}
\caption{Process of wall formation in a system $50\times8\times8$ with $\nu = 0.08$. In (a), (c), and (e), the circles indicate the
positions of solitons. In  Fig. (b), (d), and (f), the circles indicate the positions of the reversed spins.
        (a) and (b) $T=2.1 J_{\perp}$, no walls.
        (c) and (d) $T=2.0 J_{\perp}$, 2 soliton walls.
        (e) and (f) $T=1.6 J_{\perp}$, 4 soliton walls}
\label{figWalls}
\end{figure}

Here we consider a system with a size $50 \times 8 \times 8$ sites and a concentration of neutral solitons $\nu = 0.08$.

To demonstrate explicitly the formation and multiplication of domain walls,
we shall show the patterns of soliton density (Fig.\ref{figWalls}a,c,e), which is of our direct interest, and also the patterns of the reversed spins of the effective Ising model (Fig.\ref{figWalls}b,d,f) which give a useful complementary insight. Here and below, the sites with the major
orientation of spins will be left as blank space, while the sites with reversed spins will be marked in black. The edges of black areas in the
longitudinal (chains') direction indicate the positions of kinks.

At higher temperatures ($T=2.1J_{\perp} > T_2$, Fig.\ref{figWalls}a,b) we observe an ordered state impregnated by the gas of bisolitons.
When the first pair of walls condenses (at $T=2.0J_{\perp} \approx T_2$, Fig.\ref{figWalls}c,d), the concentration of noncondensed solitons
drops, then with decreasing temperature it further decreases gradually, to cure the defects in already existing walls, then it drops sharply again with the formation of the next pair of walls ($T=1.6 J_{\perp} < T_2$, Fig.\ref{figWalls}e,f). Meanwhile, the initial pair of walls diverges loosing the mutual correlation and opening the whole domain of reversed spins in between.
Comparing the distributions of Ising spins and solitons at different temperatures, we see that the number of reversed Ising spins is not conserved
(which means that the Ising magnetization can drastically drop below $T_2$ and take any value between $0$ and $1-\nu$), whereas the number of
solitons is preserved.

Interestingly, the second pair of walls nucleates in the vicinity of the first one: the incipient second pair of walls is more stable there. It
happens because the movements of solitons intending to build the second wall are prohibited towards the first one, which twice reduces their
escape probability, hence promoting the aggregation. For systems with smaller sizes, we can perform long enough simulations, when this transient
effect vanishes. However, we believe that since our elementary soliton movements are chosen in a natural way, similar to a real time soliton
dynamics, then this effect of correlated emergence of walls can take place in real systems.

\subsubsection{Integrated characteristics for the neutral system}

The obtained patterns and their evolution show up also in integrated characteristics, which are more accessible for measuring. Thus, we have
calculated the temperature dependencies of the Ising spin magnetization and of the number of transverse neighbors.
Here we consider $3D$ systems with sizes $50\times8\times8$ and $100 \times 20 \times 20$ sites for the concentration of neutral solitons $\nu =
0.08$.

\begin{figure}[tbh]
\includegraphics[width=1.0\linewidth]{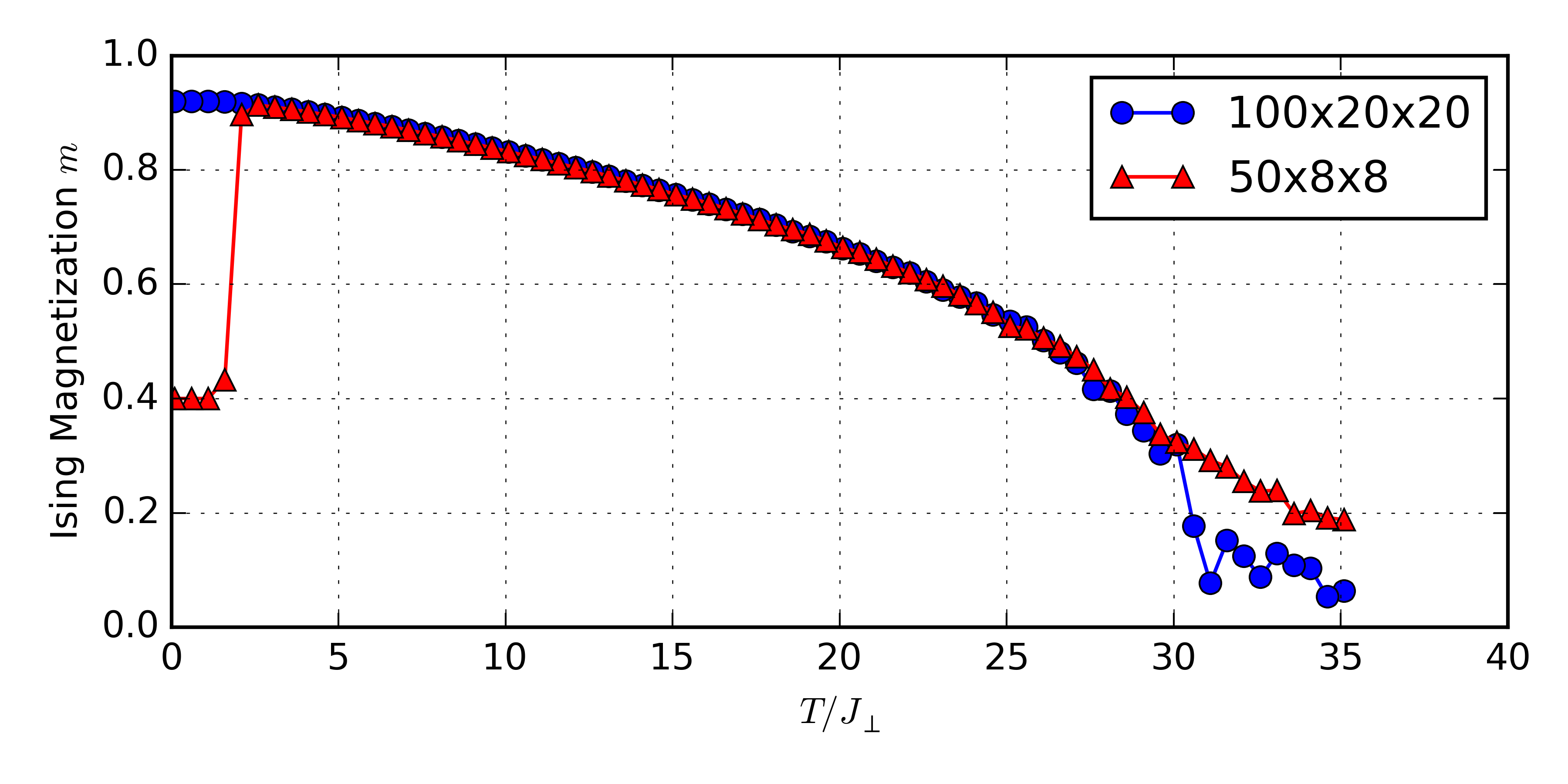}%
\centering
\caption{(Color online) Plot of magnetization $m$ of Ising spins vs temperature (in units of $J_{\perp}$) for systems with sizes $100\times20\times20$ (circles) and
$50\times8\times8$ (triangles), with $\nu = 0.08$.}
\label{plotMvsT}
\end{figure}

First, consider the Ising magnetization $m(T)$ dependence, which is shown in Fig.\ref{plotMvsT}. Excluding the low-temperature region, this dependence is very similar to the standard
plot for the Ising model with the transition temperature
$T_1 \approx 30 J_{\perp}$. The transition is smeared due to the finite-size effects.

Apparently, even this not very small fixed concentration of solitons does not affect much the high-$T$ properties, unlike the drastic effect we
see at low $T$. The $T_2$ phase transition happens with the dimensionality reduction of the system. From the thermodynamical point of view, when
the interaction between layers vanishes ($J_{||} = 0$), the magnetization must drop to $0$.
Below we explain why it may not drop to $0$ in a numerical simulation.

For the smaller system $50\times8\times8$ with lowering temperature, we see a sharp drop in the $m(T)$ dependence at $T\approx 2J_{\perp}$.
This reflects the walls formation transition at $T_2$: as it was explained in Sec. \ref{Tentative_phase_diagram} and was demonstrated
explicitly in Sec. \ref{Condensation_of_solitons_into_walls_for_neutral_solitons}, the bisolitons aggregate into walls, then these bisoliton
walls divide into single-soliton ones, and the Ising spin magnetization drops. Depending on a numerical experiment, $m$ picks a random value
between $0$ and $1-\nu$ when the system freezes at $T=0$, which is a finite-size effect associated with the finite length of the sample $L$. For a sample of a macroscopic length, however, it must be $m(0)=0$ since spin-up and
spin-down domains are equally probable.

For the bigger system $100\times20\times20$, we have observed that the walls appear in pairs, with only one reversed spin per chain, and a long
time is necessary for the walls to diverge, opening a growing domain of reversed spins in between. Typical times of the simulation are not big
enough to observe the wall splitting. This happens because in order for the bisoliton wall to divide, the system must pass through an
energetically unfavorable state: when the incipient layer of new domain grows to a disk of radius $r$, the energy increases by $2\pi r \cdot
2J_{\perp}$. If the transverse size $H$ is big enough, this high energy state cannot be reached during the time of the simulation, and the system
cannot skip from one energy minimum to another with the shifted wall. Therefore for such big systems only bisoliton walls are observed and the
Ising magnetization does not decrease at low temperatures.

\begin{figure}[tbh]
\includegraphics[width=1.0\linewidth]{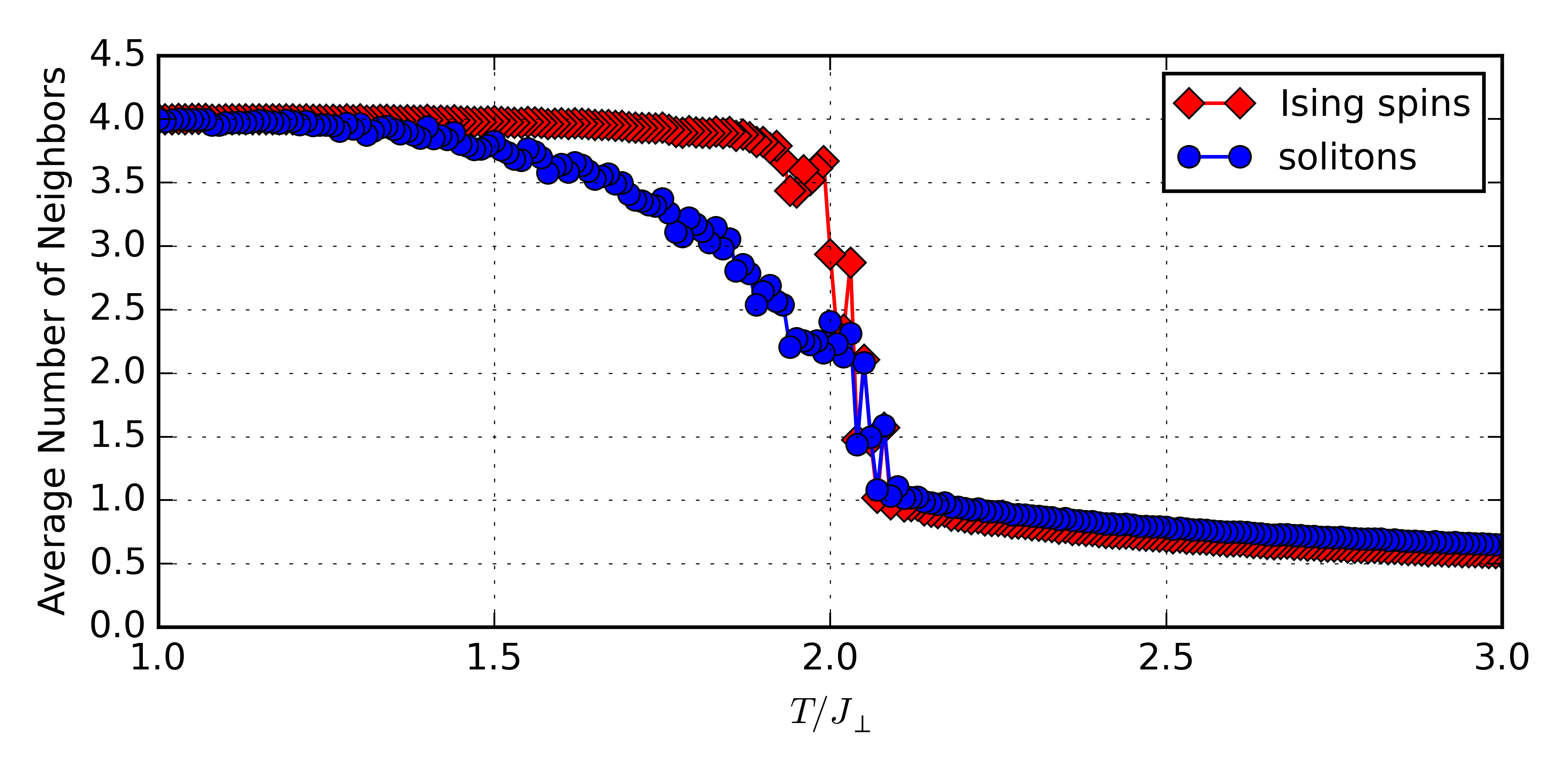}%
\centering
\caption{(Color online) Average number of transverse neighbors vs temperature (in units of $J_{\perp}$) in a system $50\times8\times8$ with $\nu = 0.08$; for the reversed spins it is shown with squares, for solitons it is shown with circles.}
\label{plotNeighbvsT8x8}
\end{figure}

Second, we consider another integrated characteristics -- the average number of transverse neighbors, which describes the transverse correlations in the system. It is particularly interesting to do so near $T=T_2$.
Since $T_2 \ll T_1$, then in the vicinity of $T_2$ only a relatively small number of reversed spins is left, being dispersed within the major
domain of aligned spins.

Thus, in order to characterize the degree of aggregation, we calculate for each reversed Ising spin (or each soliton)
the number of its neighboring reversed spins (or neighboring solitons) in the transverse direction.
Since for the domain-wall phase the number of bulk spins is much greater than the number of spins at the interfaces, then the number of reversed
spins' neighbors must jump to approximately $4$ at $T=T_2$. However for solitons, when $T$ decreases below $T_2$, a substantial number of them also condenses into walls, but this number is comparable with the number of the noncondensed solitons, therefore the jump must be not that big as for the Ising spins.

This reasoning is confirmed by Fig.\ref{plotNeighbvsT8x8}, which shows the temperature dependence of the average number of neighbors for Ising spins
and for solitons. With lowering temperature, at $T \simeq 2.0 J_{\perp}$, we see a sudden jump of the number of the reversed spins' neighbors from $\sim 1$
to $\sim 4$, which indicates the formation of domain walls as it is explicitly confirmed in Fig.\ref{figWalls}. The observed transition
temperature is in good agreement with the theoretical prediction (\ref{T2-3D}): $T_2 \approx 1.98 J_{\perp}$. The corresponding plot for
solitons also shows a jump, however, not that big: from $\sim 1$ to $\sim 2$ average number of neighbors. At $T>T_2$, the two plots essentially
coincide since almost all bisoliton pairs have the minimal size of $1$ site.

\subsubsection{Intermediate Coulomb interaction}

Now we consider the case of electrically charged solitons for a system of $50\times 8\times8$ sites with $\nu = 0.08$.
The weak CI (according to estimates, given in the Appendix B) with the Coulomb parameter $V_C = e^2/\epsilon a_{\perp} \lesssim J_{\perp}/H^2$
does not affect the system qualitatively. For example, for $V_C = 0.01 J_{\perp}$, it only lowers the temperature of condensation of solitons into walls
by $10\%$ down to $T_2'\sim 1.8J_{\perp}$

Therefore, in this section, we focus on a more interesting case of intermediate values of the Coulomb parameter $J_{\perp}/H^2 \ll V_C \ll
J_{\perp}$. In this case, the Coulomb interaction is weak enough locally, so it does not prevent the binding of solitons into bisolitons, and even
does not destroy the initial correlation of bisolitons at neighboring chains. However, it affects the large-scale structures such as domain
walls, since for $V_C \gg J_{\perp}/H^2$ the wall formation becomes energetically unfavorable.

\begin{figure}[tbh]
\centering
\subfloat[$V_C=0.02 J_{\perp}$ \label{figWallsCoulomb_a}]{%
  \includegraphics[,width=1.0\linewidth]{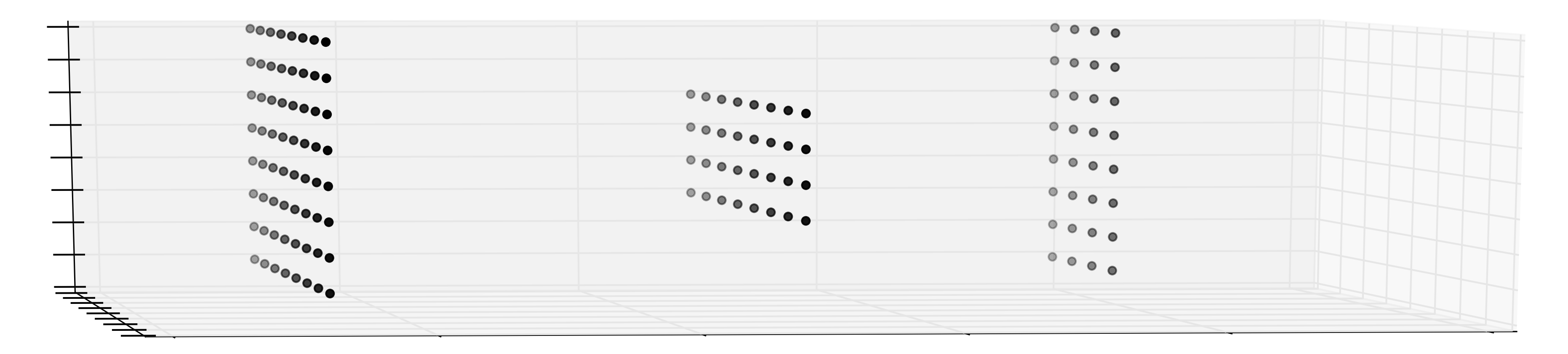}%
}

\subfloat[$V_C=0.1 J_{\perp}$ \label{figWallsCoulomb_c}]{%
  \includegraphics[width=1.0\linewidth]{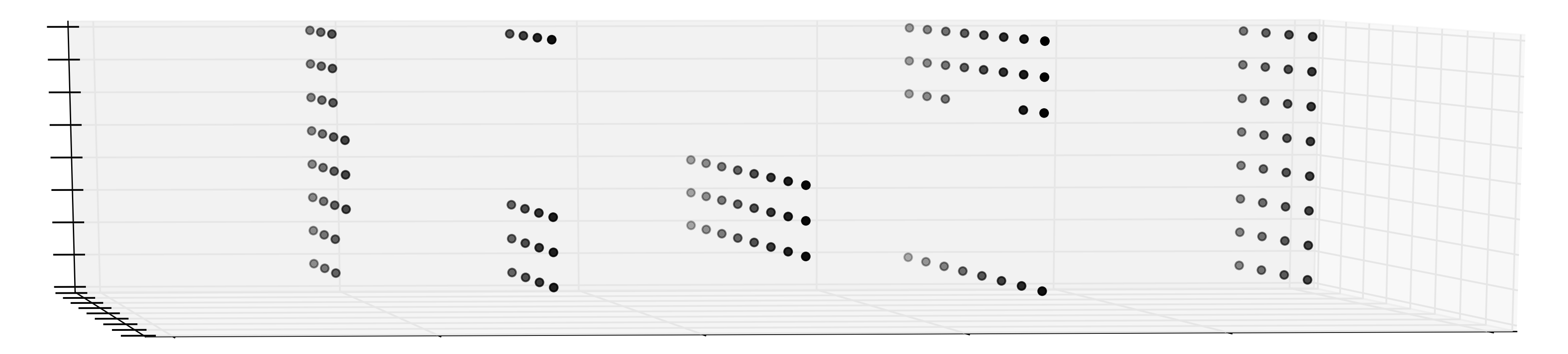}%
}

\subfloat[$V_C=0.3 J_{\perp}$ \label{figWallsCoulomb_d}]{%
  \includegraphics[width=1.0\linewidth]{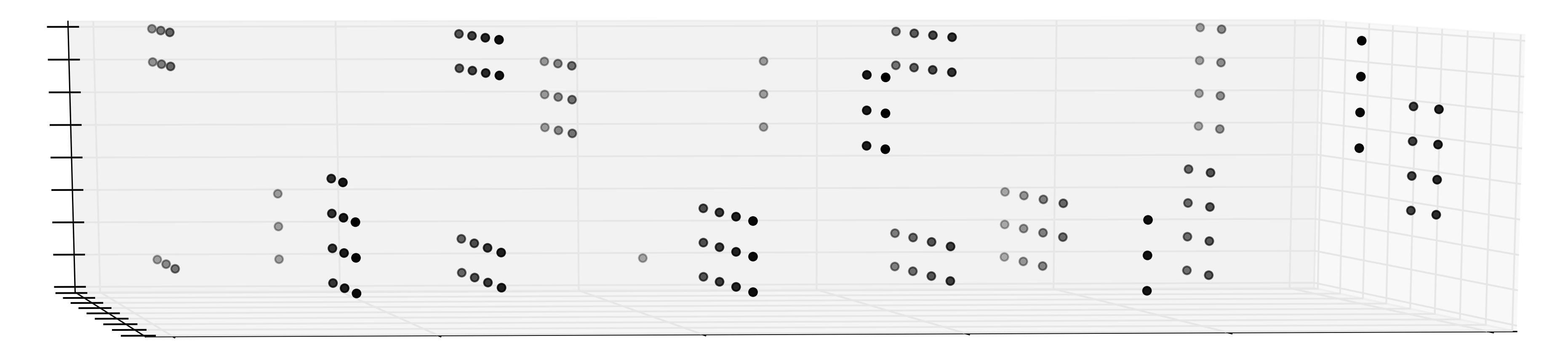}%
}
\caption{Disintegration of domain walls as CI increases. Ising spin representation of a system $50\times8\times8$ with $\nu = 0.08$ at $T=0.1
J_{\perp}$ for different values of the Coulomb parameter.  The closed dots indicate the positions of the reversed spins.}
\label{figWallsCoulombWithDefects}
\end{figure}

In contrast to the cases of neutral solitons (Fig.\ref{plotNeighbvsT8x8}) and weak CI, we do not observe sharp wall formation transition for
the intermediate values of the CI -- the temperature dependence of the average number of neighbors does not show any jumps. When $V_C$ becomes larger than
$J_{\perp}/H^2$, we still observe bisoliton walls at nonzero temperature, which now have defects (holes). As $T$ goes to $0$, these defects are grouping and can cut a wall along one of the transverse
directions (Fig.\ref{figWallsCoulombWithDefects}a).

As $V_C$ further increases, we do not observe plane walls, but only filamentary stripes, which are infinite along one transverse direction and finite along the
other (Fig.\ref{figWallsCoulombWithDefects}b, recall that periodical boundary conditions are imposed). However, it is clear that an infinite system cannot possess such infinite stripes, because they
are inefficient in terms of the interchain energy $J_{\perp}$.

Our interpretation for formation of lines rather than planes is that here the system demonstrates an extreme sensitivity of CIs to the
transverse finite-size geometry of the sample. A charged wall would create a constant electric field $E=4\pi e/(\epsilon_{||}
a_{\bot}^2)$ in the chains' $x$ direction, whose repulsive force $eE$ would oppose directly the attractive confinement force $J_{\perp}$ overpassing it at the
intermediate CI, hence no stable planes could exist. However, forming the stripes (in one transverse $y$ direction) at the expense of the part of the
confinement energy (lost in the other transverse direction $z$), the system generates a decreasing electric field $E\propto 1/\sqrt{x^2+z^2}$
which falls below the confinement force at sufficiently large distances, hence preserving this partial aggregation.

Therefore, in finite systems, the disintegration of plane walls happens via formation of linear structures. For infinite systems, we expect
that these structures will further disintegrate into finite disks (similar to described in the next paragraph ones).

If we choose higher value of the Coulomb parameter, the formation of stripes even for a finite system becomes also energetically unfavorable
and we observe only disks of bisolitons (Fig.\ref{figWallsCoulombWithDefects}c). Observed disklike formations are consistent with
the analytical results presented in Appendix B, where it is shown that the maximum radius of the disks is $R^*\sim \sqrt{J_{\perp}/V_C}$.

\subsubsection{Strong Coulomb interaction}

\begin{figure}[tbh]
\centering
\subfloat[$V_C=10 J_{\perp}$ \label{1}]{%
  \includegraphics[,width=1.0\linewidth]{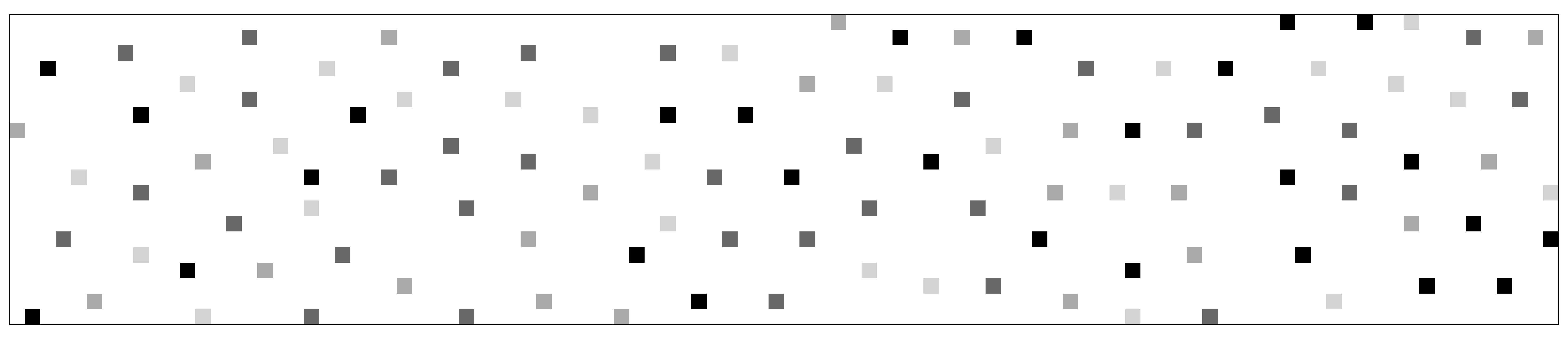}%
}

\subfloat[$V_C=100 J_{\perp}$ \label{2}]{%
  \includegraphics[width=1.0\linewidth]{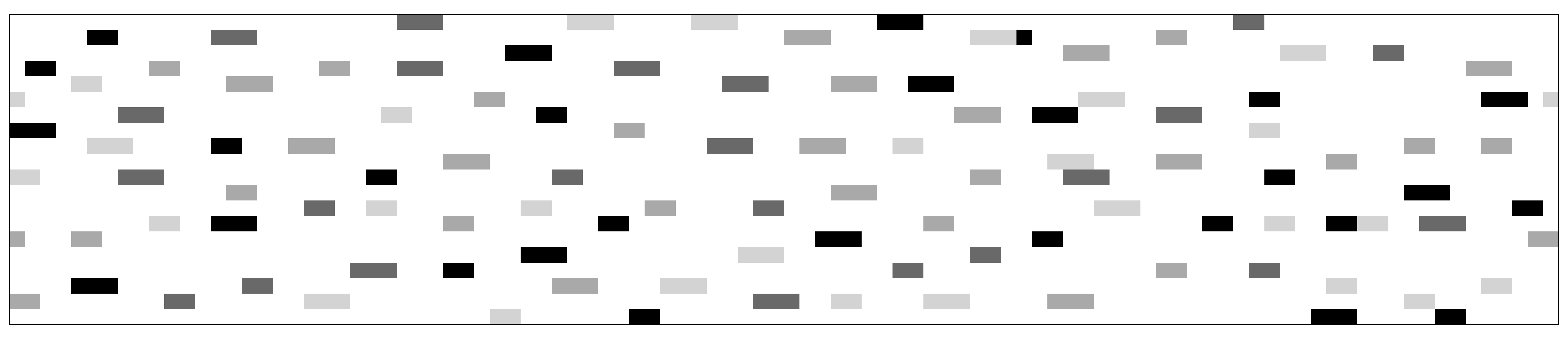}%
}
\caption{Ising spin representation of four neighboring $100\times20$ slices of a $100\times20\times20$ system, which are projected onto the $xy$ plane. Reversed spins from four projected planes are marked in different shades of gray.
For $V_C = 10J_{\perp}$ (a), we observe a "liquid" of the reversed Ising spins (bisolitons at the minimal distance); for $V_C = 100 J_{\perp}$ (b), we see that the bisoliton size increases (here $T=0.1 J_{\perp}$ for both cases).}
\label{figHighV1}
\end{figure}

Here we consider a system of $100\times 20\times20$ sites with $\nu = 0.026$ (the parameters were chosen in order to compensate incommensurateness effects, as explained below).
Strong CIs $V_C \gtrsim J_{\perp}$ affect the local bisoliton pairing.
The transverse disks shrink to the minimum size of $1$ bisoliton.
When $V_C$ further increases, these bisolitons form a Wigner "liquid" (Fig.\ref{figHighV1}a)
(with a short-range order of individual solitons -- contrarily to a Wigner crystal with a long-range order). It is known that the ground-state
distribution of charges must form a triangular lattice in $2D$ [\onlinecite{Bonsall:1977}] and a body-centered cubic lattice in $3D$
[\onlinecite{Drummond:2004}]. However, for systems with finite discretization, the commensurateness effects become very important: even small
incommensurateness destroys the long-range order, while the short-range order persists \cite{Aubry:1983}.

\begin{figure}[tbh]
\centering
\subfloat[Ising spins\label{1}]{%
  \includegraphics[width=1.0\linewidth]{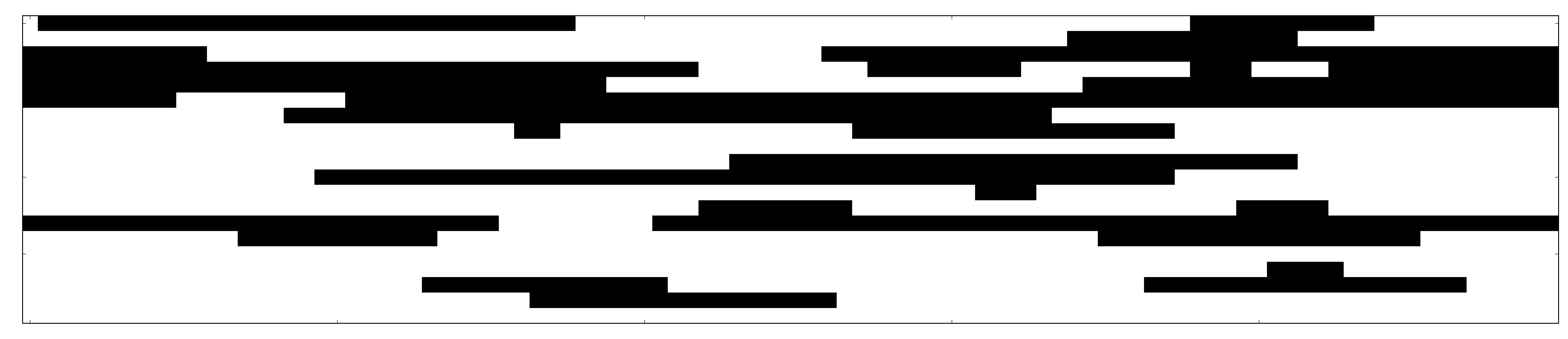}%
}

\subfloat[Solitons\label{2}]{%
  \includegraphics[width=1.0\linewidth]{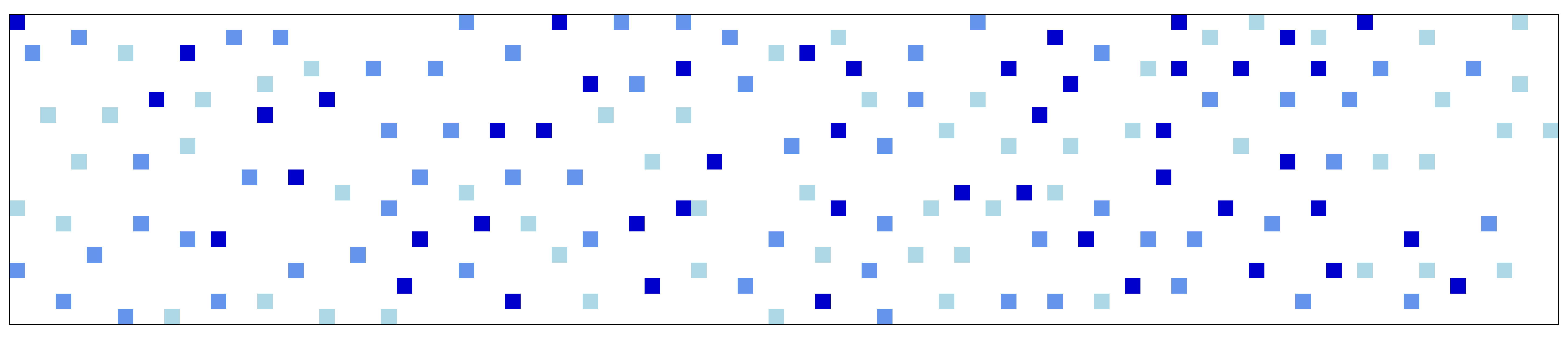}%
}
\caption{(Color online) System $100\times20\times20$ for $V_C = 1000 J_{\perp}$, $T=0.1 J_{\perp}$. (a) Ising spin representation of $100\times20$ slice of the system. The Ising order is destroyed. (b) Soliton representation of $100\times20\times3$ slice, which is projected on $xy$ plane, solitons from different planes are marked in different shades. The solitons are deconfined,  a "liquid" of an individual solitons rather than bisolitons is observed.}
\label{figHighV2}
\end{figure}

For even higher $V_C$, the Coulomb force starts to compete locally with the confinement force and the size of a bisoliton starts to grow
(Fig.\ref{figHighV1}b). Neglecting the interaction between bisolitons we can estimate their size. A bisoliton elongates until the Coulomb force is
balanced by the confinement force: $V_C/l_{bs}^2 \sim 8 J_{\perp}$, therefore $l_{bs} \sim \sqrt{V_C/8 J_{\perp}}$. This estimate holds as long
as the average bisolitons' size is much less than the distance between them: $l_{bs} \ll \nu_{bs}^{-1/3}$. When $l_{bs} \sim \nu^{-1/3}$ (which
happens at $V_C \sim 8 J_{\perp} \nu^{-2/3} \sim T_1 \nu^{1/3}$), this size becomes comparable to the distance between solitons, interactions
between them become important. Increasing the CI to the highest values, we observe that the Ising order is destroyed in favor of a Wigner
"liquid" of individual solitons, rather than bisolitons (Fig.\ref{figHighV2}a,b).

\subsection{Numerical results for the 2D case}

\subsubsection{System of neutral solitons}

In this section, we consider a $2D$ system with the size of $200 \times 25$ sites and concentration of neural solitons $\nu = 0.03$. As discussed
in the Sec. \ref{Tentative_phase_diagram}, well below the Ising transition temperature $T_1$ (for the considered system $T_1 \approx 20
J_{\perp}$), there exists a characteristic temperature $T_2$ at which perpendicular rods start to form with their characteristic length gradually
increasing with lowering $T$. For our finite samples, there is also a width $H$ dependent temperature $T_F(H)$ at which these rods become long
enough to pass across the entire sample.

\begin{figure}[tbh]
\centering

\subfloat[$T=28 J_{\perp}$  \label{a}]{%
  \includegraphics[width=1.0\linewidth]{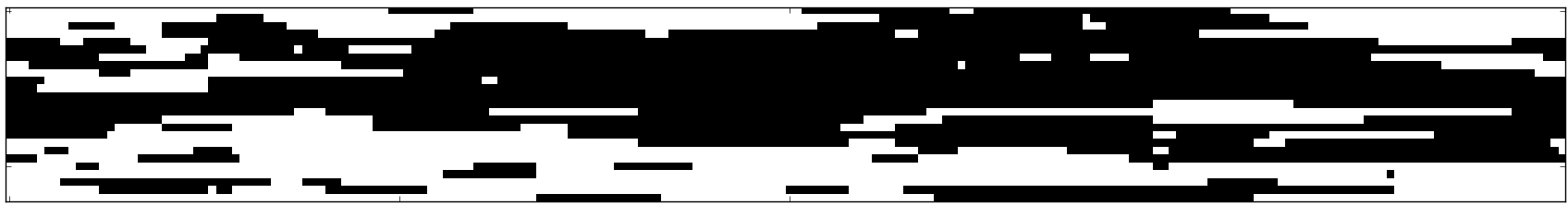}% folder: 0
}

\subfloat[$T=10 J_{\perp}$  \label{a}]{%
  \includegraphics[width=1.0\linewidth]{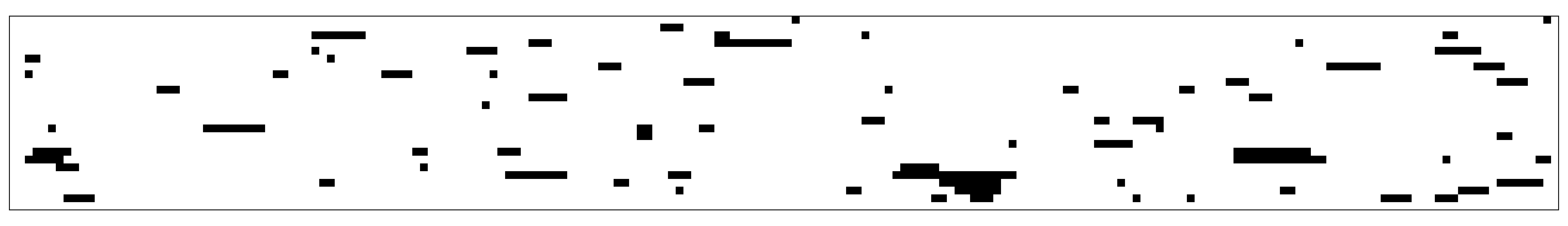}% 0c
}

\subfloat[$T=1.5 J_{\perp}$  \label{b}]{%
  \includegraphics[width=1.0\linewidth]{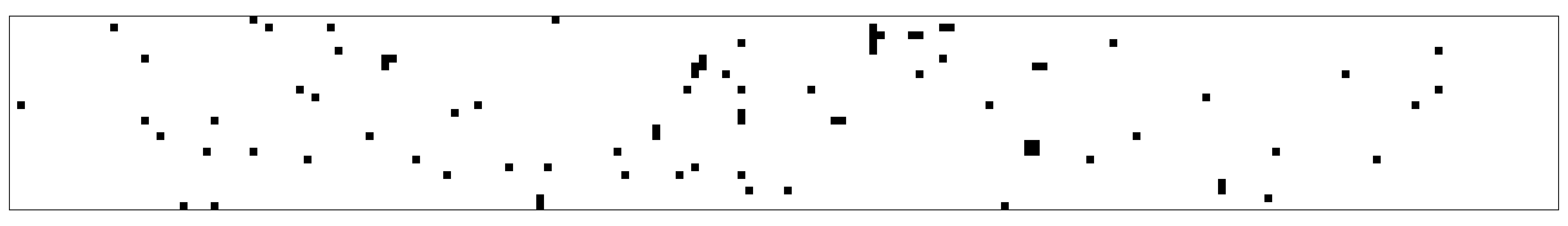}% 1d
}

\subfloat[$T=0.8 J_{\perp}$ \label{c}]{%
  \includegraphics[width=1.0\linewidth]{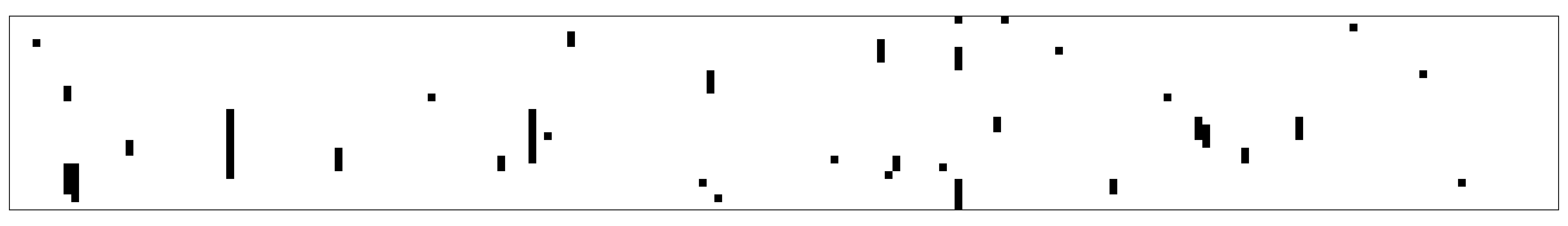}% 1d
}

\subfloat[$T=0.7 J_{\perp}$  \label{b}]{%
  \includegraphics[width=1.0\linewidth]{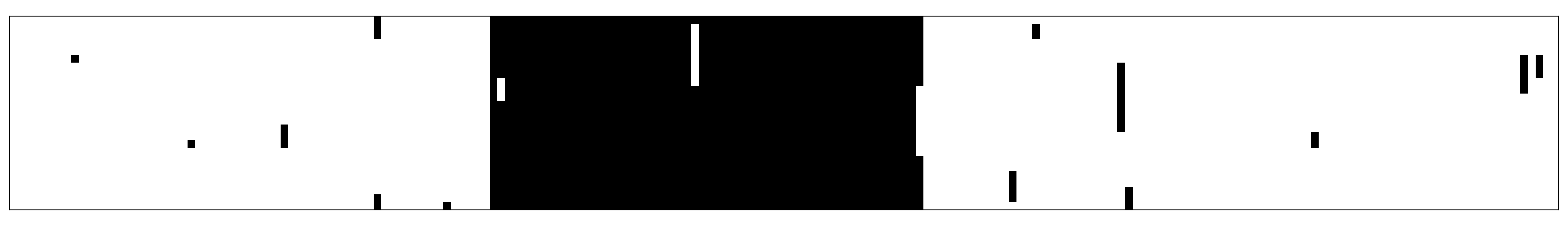}% 1d
}
\caption{Neutral solitons: the Ising spin representation of a $2D$ system with $200\times25$ sites and $\nu=0.03$  for different temperatures.}
\label{Ising2D}
\end{figure}

Figure \ref{Ising2D} shows the Ising spin representation of the system's evolution with lowering $T$.
For $T=28 J_{\perp} > T_1$, the Ising disordered phase is observed (Fig.\ref{Ising2D}a).
At $T=10 J_{\perp} < T_1$, we observe the Ising ordered phase with bound pairs of bikinks (Fig.\ref{Ising2D}b). These
pairs shrink when $T$ lowers (Fig.\ref{Ising2D}c). Then the rods of reversed spins change the predominant orientation from the longitudinal
one at high temperatures to the transverse one at low $T$ (Fig.\ref{Ising2D}d). The first case indicates the regime of loosely bound on-chain
pairs of kinks, while the second case indicates the transverse aggregation of tightly bound kinks. Finally, at the lowest considered $T$
(Fig.\ref{Ising2D}e), the aggregated rods cross the entire sample and domains are created.

\begin{figure}[tbh]
\centering
  \includegraphics[width=1.0\linewidth]{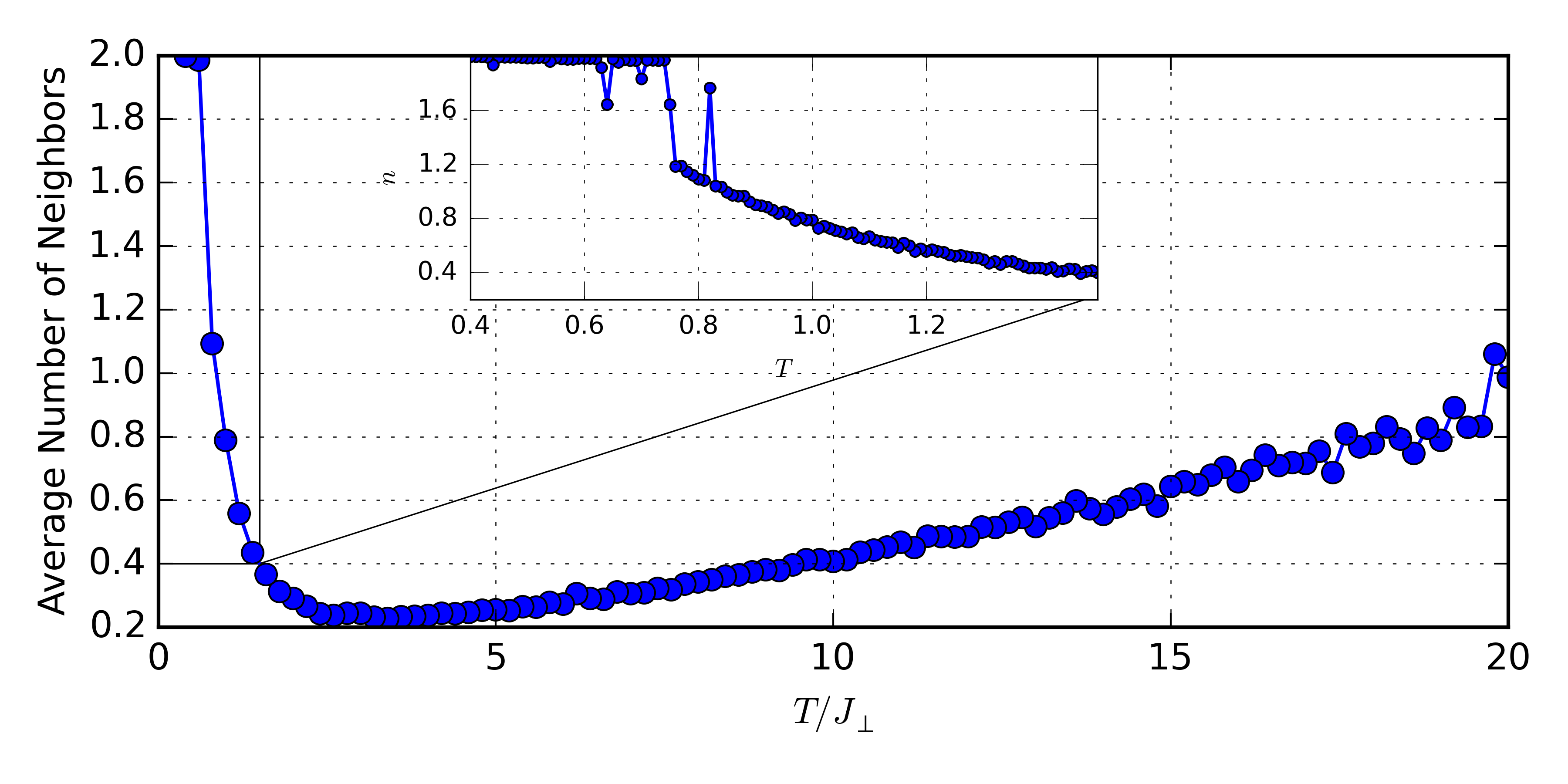}%
\caption{(Color online) Neutral solitons: average number of the reversed spins' neighbors vs temperature in a system with $200\times25$ sites and  $\nu = 0.03$. The inset shows
temperature range $T=0.4J_{\perp}..1.5J_{\perp}$}
\label{plotNeighbvsT200x25nu003}
\end{figure}

As in Sec. \ref{Numerical_results_for_3D_case}, $T_F$ lies deeply below the Ising transition temperature, only a relatively small number of
reversed spins is left, being dispersed within the
major domain of aligned spins; therefore  we characterize the degree of aggregation, by calculating for each soliton (or each
Ising spin) the number of its interchain neighbors.
For the domain wall phase the number of reversed spins' neighbors must be approximately 2.

\begin{figure}[tbh]
\centering
  \includegraphics[width=1.0\linewidth]{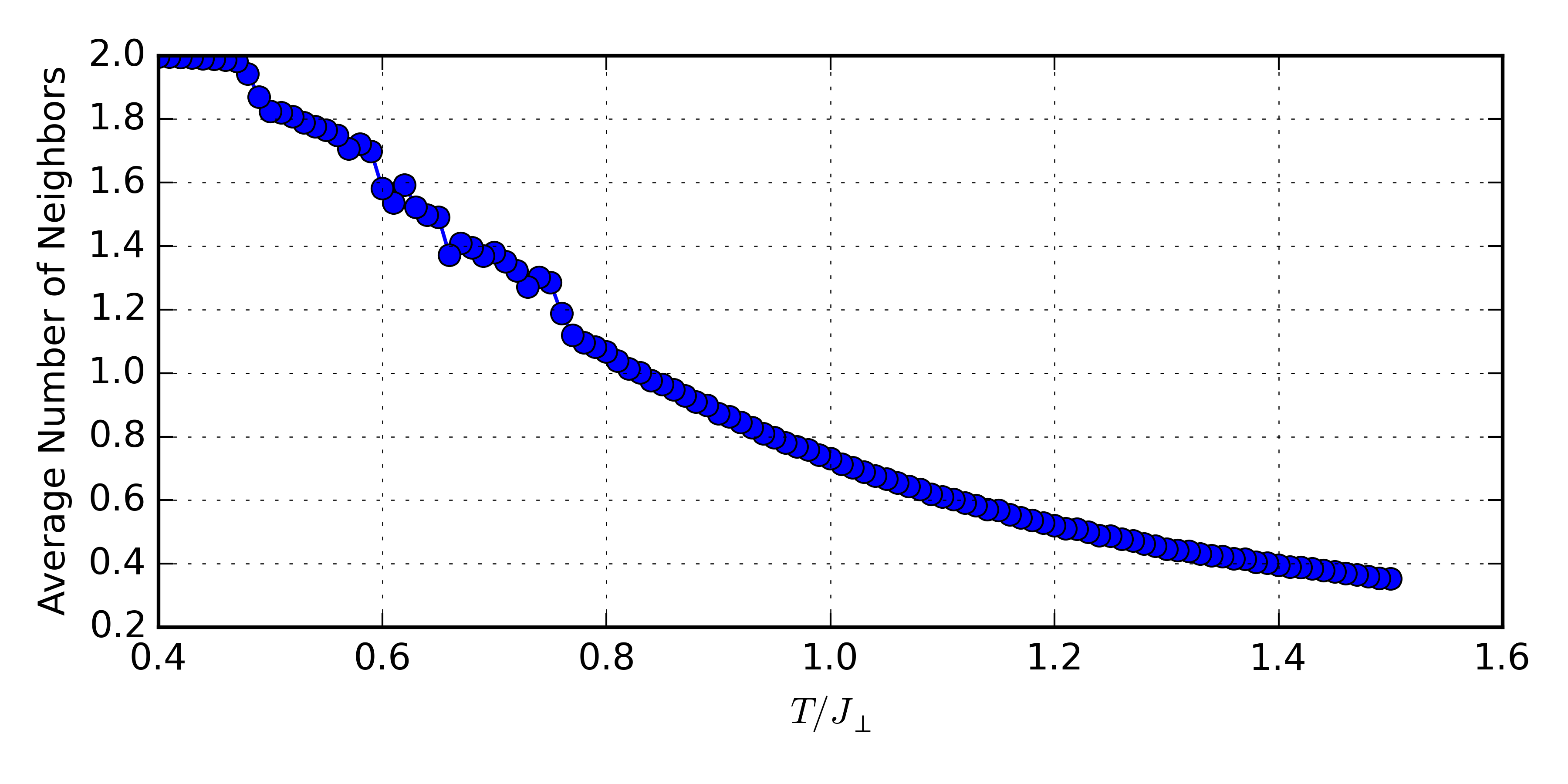}%
\caption{(Color online) Neutral solitons: the average number of soliton interchain neighbors vs temperature, in a system with $200\times25$ sites and  $\nu = 0.03$.}
\label{plotNeighbSolvsT200x25nu003}
\end{figure}

Figure \ref{plotNeighbvsT200x25nu003} shows the temperature dependence of the average number of neighbors of the reversed Ising spins. It shows that
upon cooling from the Ising transition temperature ($T_1 \approx 20 J_{\perp}$), we arrive firstly at some characteristic temperature ($T_2
\approx 2.5 J_{\perp}$) when the average number of neighbors starts to grow. On further cooling, another characteristic temperature ($T_F \approx
0.75 J_{\perp}$) is reached when the average number of neighbors suddenly changes from $\sim 1.2$ to $\sim 2$, which indicates the wall
formation process.
Figure \ref{plotNeighbSolvsT200x25nu003} shows the temperature dependence of the average number of soliton interchain neighbors.
We see that their number increases gradually with lowering temperature and shows no peculiarities. This means that the transition at $T_F$ is due
to finite-size effects and in an infinite sample we shall observe the condensation of solitons into infinite lines only at $T=0$.

\subsubsection{System of charged solitons}

\begin{figure}[tbh]
\centering
\includegraphics[width=1.0\linewidth]{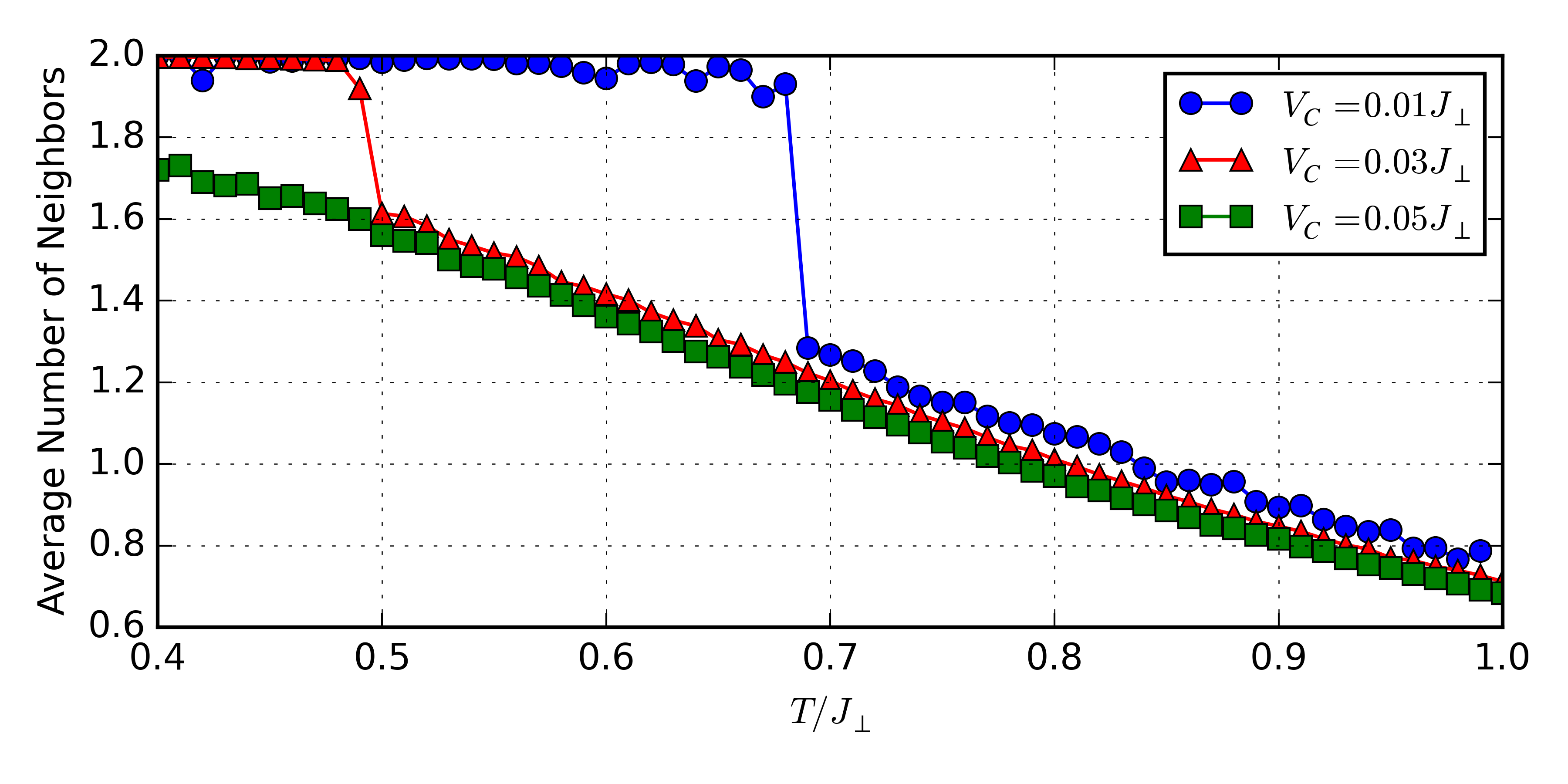}%
\caption{(Color online) Charged solitons: average number of reversed spins' neighbors vs temperature in a system with $200\times25$ sites and $\nu = 0.03$ for
$V_C=0.01J_{\perp}$ (circles), $V_C=0.03J_{\perp}$ (triangles), and $V_C=0.05J_{\perp}$ (squares).}
\label{plotNeighbvsT200x25nu003_V}
\end{figure}

Here we consider a similar system, but now with charged solitons. For  $V_C = 0.01 J_{\perp}$ (Fig.\ref{plotNeighbvsT200x25nu003_V}), the
temperature $T_F$ is lowered with respect to the case $V_C = 0$ (Fig.\ref{plotNeighbvsT200x25nu003}) -- now  the wall formation transition is
observed at the lower $T'_F \approx 0.69 J_{\perp}$. Increasing the Coulomb parameter up to $V_C = 0.03 J_{\perp}$
(Fig.\ref{plotNeighbvsT200x25nu003_V}), we see that the wall formation transition temperature is further lowered down to $T''_F \approx 0.5
J_{\perp}$.

With increasing $V_C$, $T_F$ eventually decreases to $0$. For some critical value of the Coulomb parameter, we get $T_F=0$, in which case rods grow
up only  to a maximum size $l^* \sim J_{\perp}/V_C \gg 1$ (see Appendix B for the details of analytical estimations). However, precise observation of this critical value of $V_C$ is difficult, since the acceptance rate of the algorithm exponentially vanishes to $0$ as $T \rightarrow 0$.

For strong CI $V_C \gtrsim 1$, the behavior of the $2D$ system is qualitatively the same as for $3D$. The transverse rods shrink to the minimum
size of $1$ bisoliton, then the CIs start to compete with the confinement force, so bisolitons start to elongate and, at some $V_C$, the Ising
order is destroyed. For the highest values of CIs a Wigner "liquid" of individual solitons is observed, which case was studied in \cite{Lee:2001,Lee:2002,Zaanen:2013}.

\section{Discussion and conclusions.}

We have presented numerical  and qualitative analysis of phase transitions in ensembles of solitons as they can be created and studied in
experiments on optical pumping and field effect doping in systems with cooperative electronic states.

For $3D$ systems of neutral solitons, as temperature lowers, we observe two phase transitions. The first transition at $T_1$ reflects the spin
ordering of the equivalent Ising model. In terms of the original solitons, $T_1$ is the temperature, below which individual solitons become
confined into bisoliton pairs. With further decreasing temperature, the size of a pair decreases, reaching the minimal value of $1$ at $T\sim
J_{\perp}$, when a gas of bisolitons forms. With further cooling of the system, the bisolitons start to aggregate into transverse disklike formations. Finally, at some critical temperature $T_2$, the second phase transition occurs: these disks cross the entire sample, domain walls are
formed, and the Ising magnetization drops to $0$. The dimensionality of the system effectively reduces to $D=2$.

For $3D$ systems of charged solitons, the locally small CI (when $V_C \ll J_{\perp}$) can nevertheless affect the $T_2$ transition, where
macroscopic patterns are created. This happens because a large scale structure, such as a domain wall, gives rise to a high long-range electric
field, which erases the gain of the confinement energy reached by the wall formation.
For a macroscopic system without an external screening, even an arbitrary small Coulomb parameter $V_C \neq 0$ destroys the walls, only disklike
formations with the maximum size $R^* \propto \sqrt{J_{\perp}/V_C}$ are observed. However, if the screening is present and the screening
length $l_s < R^*$, then these domain walls still cross the entire sample (in our numerical study the sample width $H$ plays a role of the
screening length $l_s$).

For high values of the CI, the disklike formations disintegrate into separate bisolitons. When the CI is further increased, bisolitons arrange
in a Wigner liquid state. Further, bisolitons start to elongate and when their size becomes comparable with the interpair distance,
the Ising order breaks and a Wigner liquid of individual solitons is observed.

Neutral $2D$ systems behave qualitatively similar to the $3D$ case  at high and intermediate temperatures, the Ising-like transition at $T_1$ still
exists. However, an important distinction is that $T_2$ is rather a crossover temperature in $2D$: the growing rods do not cross the entire
sample for a macroscopic system. However, for a finite one, there still exists some temperature $T_F$ of domain wall formation. For charged $2D$
system, $T_F$ lowers with the increase of the CI and, when it reaches $0$, only rods of finite transverse length $l^* \propto J_{\perp}/V_C$
are observed even at $T=0$.

The presented results of the MC simulation for neutral solitons agree with the earlier predictions \cite{Bohr:1983} (for both $2D$ and $3D$
cases). However, the results for charged solitons do not completely agree with the previous work \cite{Teber:2001}, performed only for the $2D$
case. There it was observed that with increasing $V_C$  at $T=0$, domain walls are not destroyed by the CI, but they are rather roughened. This
difference in the results occurs presumably because in \cite{Teber:2001} the ensemble of solitons was treated with preserving number of
bisolitons at each chain, rather than only globally. This limitation is overcome in our treatment which also has employed a more efficient
algorithm allowing for modeling of $3D$ systems, even for charged particles. This approach matches the condition of relaxation of a soliton
system after a fast optical pumping or an impact of a strong electric field, considered in the present work: only bisolitons can jump between the
chains. Actually, there are electron pairs that jump while the order parameter is adjusted to their presence or absence.

For the experiments with optical pumping to the gas of solitons, we predict that along the equilibration trajectory the system will experience two
phase transitions: confinement of solitons at high $T$ and their aggregation to the stripe phase at lower $T$.

The presented theoretical picture should find its experimental realization. There can be traditional, already verified methods of solitons’ identifications (recall the discussion and references in Sec.\ref{New_accesses_to microscopic_solitons_in_q1D_electronic_systems}). Not all of them, particularly the latest and most spectacular direct visualizations by the STM, can be applied in conditions of pump- or field-induced experiments, but they will serve as a preliminary assurance for the correctly chosen system. Among traditional techniques, the pump-and-probe optics can trace the time evolution of solitons via their associated spectral features; that have been already so efficiently exploited in conducting polymers \cite{vardeny,spin-kinks}.  In the modern, and closer to our discussion, context of the fast optical pumping with optical probes, the solitons appeared already in studies of neutral-ionic transitions \cite{NI-solitons,Okamoto-NI, Uemura:2010, Miyamoto:2012}.
The state of an ensemble of solitons, particularly the aggregation into regular stripes, may be traced by methods of the “time-sliced diffraction”. Recently, the time resolution of the electron- and x-ray diffraction (see the latest publications  \cite{Huber:2014, Gulde:2014, Haupt:2016}  and references therein) has been pushed to the subpicosecond scale, which makes it very promising to study the induced phase transitions accompanied by the structural aggregation.

We mention finally that the studied here phase transitions with confinement and with segregation in the ensemble of topologically nontrivial excitations may serve as elementary illustrations to unresolved yet issues in high-energy physics (confinement of quarks) and cosmology (phase transitions in the early universe).

%\section*{Acknowledgements}
%We acknowledge the financial support of the Ministry of Education and Science of the Russian Federation in the framework of Increase Competitiveness Program of NUST “MISiS” (No K2-2014-015).

%__________________________________________________________________
%                         Appendix A.
\appendix{}

\section{Analytical results for a neutral system}

\subsection{$2D$ case}
Consider the high temperature regime $T \gtrsim T_1$. The exact expression for the solitons' concentration at the critical temperature is
\cite{Bohr:1983}
\begin{equation*}
\nu (T_1, J_{\perp}) = \frac{1}{2} - \frac{1}{\pi} \cosh \frac{2J_{\perp}}{T_1} \arctan \frac{1}{\sinh 2J_{\perp}/T_1},
%\label{AppA_nu1}
\end{equation*}
then for the case $\nu \ll 1$, we get
\begin{equation}
T_{1}\approx \frac{2J_{\perp}}{\pi \nu}.
\label{Appa_T1}
\end{equation}
Now consider the low temperature regime $T \ll J_{\perp}$.
In this limit all bisoliton pairs are shrunk to the minimal size of one reversed spin, then the magnetization is $m=1-2\nu_{bs}=1-\nu$. Using the
exact expression for magnetization in the $2D$ Ising model \cite{McCoy:1973}:
\begin{equation}
m_{2D}(T,J_{\perp},J_{||}) = \left[ 1- \left( \sinh \frac{2J_{\perp}}{T} \sinh \frac{2J_{||}}{T} \right)^{-2} \right]^{1/8},
\label{m2D}
\end{equation}
we find that
\begin{align}
J_{||}(T,\nu) &= \frac{T}{2} \arcsinh \frac{1}{\sinh(2J_{\perp}/T) \sqrt{1-(1-\nu)^8}} \approx \nonumber \\
&\approx \frac{T \exp(-2J_{\perp}/T)}{2\sqrt{2\nu}}.
\label{AppA_Jpar2D}
\end{align}
Crossover from the bikinks gas state to the state of growing transverse rods occurs when $J_{||}(T,\nu)$ becomes of order of $J_{\perp}$. Then from
(\ref{AppA_Jpar2D}) for $\nu \ll 1$, we get with logarithmic accuracy
\begin{equation}
T_2 \approx \frac{4J_{\perp}}{\ln(1/\nu)}.
\label{AppA_T2}
\end{equation}

\subsection{$3D$ case}

To find an estimation for $J_{||}(T,\nu)$ at low temperatures $T \ll J_{\perp}$ , we employ an approximation where interactions in the transverse
planes are considered exactly while the weak interactions between planes are taken into account using the mean-field theory.

In this approximation, the in-chain interaction energy terms per spin $S_{n,\alpha}$ becomes
\begin{equation}
-\frac{1}{2} J_{||} S_{n,\alpha} (S_{n+1,\alpha} +  S_{n-1,\alpha}) \approx -J_{||} S_{n,\alpha} m \,\equiv - H S_{n,\alpha},
\end{equation}
where $H$ is an effective magnetic field. Therefore the magnetization for $3D$ case is simply given as the magnetization of the $2D$ Ising model
in the weak effective magnetic field:
\begin{equation}
m(T) \approx m_{2D}(T) + \chi(T) H,
\label{m3D}
\end{equation}
where $\chi(T)$ is the susceptibility of the $2D$ Ising model. Using (\ref{m2D}) and (\ref{m3D}), we get
\begin{equation}
m(T) \approx \left( 1-(\sinh 2J_{\perp}/T)^{-4} \right)^{1/8} (1+\chi(T) J_{||}).
\label{m3D1}
\end{equation}

Now we link the magnetization $m$ to the soliton concentration. Since all layers are magnetized in one direction, we introduce deviations from
mean magnetization $\delta S_{n,\alpha} = S_{n,\alpha}-m$, so that $\langle S_{n,\alpha} S_{n+1,\alpha} \rangle = m^2 + \langle \delta
S_{n,\alpha} \delta S_{n+1,\alpha} \rangle$.

If $T \rightarrow T_2$ then $J_{||} \rightarrow 0$ and $\langle \delta S_{n,\alpha} \delta S_{n+1,\alpha} \rangle \rightarrow 0$, so we can
expand the latter in powers of $J_{||}$:
\begin{equation}
\langle \delta S_{n,\alpha} \delta S_{n+1,\alpha} \rangle = \alpha(T,J_{\perp}) J_{||} + O(J_{||}^2).
\label{dSdS}
\end{equation}
Using (\ref{rho}), (\ref{m3D1}), (\ref{dSdS}) we get:
\begin{equation}
J_{||} (T,J_{\perp},\nu) = \frac{1-2\nu - \left(1-\sinh^{-4} \frac{2J_{\perp}}{T} \right)^{1/4}}
{2 \chi(T, J_{\perp})\left(1-\sinh^{-4} \frac{2J_{\perp}}{T} \right)^{1/4} - \alpha(T,J_{\perp})}.
\label{Jparexact}
\end{equation}
From the condition $J_{||} (T_2,J_{\perp},\nu) = 0$ we re-derive the result of \cite{Bohr:1983}:
\begin{equation}
T_2 = \frac{2 J_{\perp}}{\arcsinh \left( \left[ 1-(1-2\nu)^4 \right]^{-1/4}\right)} \approx \frac{8J_{\perp}}{\ln 2/\nu}.
\label{T2}
\end{equation}
To estimate $J_{||} (T,J_{\perp},\nu)$ we assume that there is no special reasons for the denominator of (\ref{Jparexact}) being small. So, up to
a factor of order of 1, we can neglect $\alpha$ in (\ref{Jparexact}).

For $\chi(T,J_{\perp})$ we use the results of \cite{Nickel:1999,Nickel:2000}, where it was shown that $\chi$ can be decomposed to a well
convergent series
\begin{equation}
\begin{split}
\chi(T) = \frac{m^2}{T} \sum_{n=1}^{\infty} \hat{\chi}^{(2n)} (T), \\
\hat{\chi}^{(2)} = \frac{(1+k_<^2) E(k_<) - (1-k_<^2)K(k_<)}{3\pi (1-k_<) (1-k_<^2)^{3/4}}.
\end{split}
%\label{SS}
\end{equation}

\noindent Here, $E$ and $K$ are the complete elliptic integrals of the first and the second kind, $k_< = (\sinh 2J_{||}/T \sinh
2J_{\perp}/T)^{-1}$. For $T \ll J_{\perp}$, $k_< \approx \exp(-4J_{\perp}/T) \ll 1$, therefore $\hat{\chi}^{(2)} \approx \tfrac{1}{4} k_<^2$, and
we estimate $\chi$ as:
\begin{equation}
\chi(T,J_{\perp}) \propto \frac{m^2 \exp(-8J_{\perp}/T)}{4T} \propto \frac{\exp(-8J_{\perp}/T)}{4T}.
\label{chi-appr}
\end{equation}
Expanding (\ref{Jparexact}) in the vicinity of $T=T_2$ and using (\ref{chi-appr}), we get the desired estimation:
\begin{align}
J_{||}(T,J_{\perp},\nu) \propto \frac{64 J_{\perp} \exp(-8J_{\perp}/T_2)}{T_2} (T-T_2) \propto \nonumber \\
\propto 4\nu \ln\dfrac{2}{\nu} \cdot (T-T_2).
\end{align}
We see that $J_{\perp}$-dependence is contained only in $T_2$ and the slope of the line depends only on $\nu$.

\section{Analytical results for a charged system.}

Here we present estimations and qualitative arguments on effects of CIs.

\subsection{$2D$ case.}

Consider first the $2D$ case. According to our modeling and following the exact
results for the neutral system \cite{Bohr:1983}, we suggest that the basic units are
still the straight lines (rods) of unseparated bikinks; their lengths $l$
will be taken as dimensionless; the physical scale is $a_{\perp}$.

The CI energy of a line of bikinks is%
\[
\frac{4e^{2}}{\epsilon a_{\perp}^{2}}\int\int\frac{dydy^{\prime}}%
{|y-y^{\prime}|}\exp\left(  -\frac{|y-y^{\prime}|}{l_{s}a_{\perp}}\right)
\approx 4   V_C l\ln\min\{l,l_{s}\}%
\]
where $V_C = e^2/\epsilon a_{\perp}$, $l_{s}a_{\perp}$ is the screening length by remnant or external carries;
it appears for $l>l_{s}$. Our assumption of intermediate CIs is that locally they are
weak, i.e., $V_C\ll J_{\perp}$, but become effective for aggregates when $V_C\ln
l^{\ast}/J_{\perp}$ is not small, i.e., $V_C \gtrsim V_{inter} = J_{\perp}/\ln l^*$ ($l^{\ast}$ is the characteristic rod length).

Because of the equilibrium between rods with respect to exchange of building
units -- the bikinks, their partial chemical potentials $\mu_{l}$ are related
as $\mu_{l}=2l\mu^*$;  here we include to the definition
of $2\mu^*$ not only the soliton energy $2E_{s}$ but also the CI energy of the
elementary pair $e^2/\epsilon a_{||}$, so that $\mu^* = \mu_s - E_s - e^2/2 \epsilon a_{||}$.
\newline The distribution of segments is%
\[
n(l)=\exp(-4\beta J_{\perp}+2\beta\mu^* l-4\beta V_C l\ln(\min\{l,l_{s}\}))
\]
The parameter $\mu^*(\nu)$ is to be determined from the
self-consistency condition
\begin{align}
\nu &=2\sum_{l}n(l)l = \nonumber \\
&= 2\exp(-4\beta J_{\perp})\sum_{l}l\exp(\beta(2\mu^* l-4 V_C l\ln(\min\{l,l_{s}\})))
\label{nu(mu)}%
\end{align}
Approaching the regime of stripe formation, the sum is determined
by large $l$, then the summation can be approximated by integration over $l$.

Consider, first, the case without screening, when the characteristic rod length
$l^{\ast}\ll l_{s}$. Now,%
\begin{equation}
\nu\exp(4\beta J_{\perp}) = 2\int dl~l\exp(\beta(2\mu^* l-4 V_C l\ln l))%
\label{int-nu}%
\end{equation}
The exponent $S(l)=\beta(2\mu^* l-4V_C l\ln(l))$ in (\ref{int-nu}) is rapidly
decreasing at large $l$ and the integral is convergent, because of the $V$
term, at any -- even positive -- $\mu^*$. For negative $\mu^*$ it decreases
rapidly at all $l$ and the sum is saturated already by $n(1)=\exp
(\beta(-4J_{\perp}+2\mu^*)))$ with $\nu\approx n(1)$. With decreasing $T$,
$\mu^*$ increases, it reaches zero with no qualitative effect unlike the case of
neutral systems, and finally becomes positive when $S(l)$ acquires the rising
part at $1<l<l^*$ where the maximum position given by $dS/dl=0$ is
$l^*\approx C\exp(\mu^{\ast}/2V_C)$, which
makes precise the definition of $l^*$. The saddle-point
approximation for the integral over $l$ gives%
\begin{align*}
& \nu \exp(4\beta J_{\perp}) \approx 2 \left( \frac{2\pi}{|d^{2}S/dl^{2}|}\right)^{1/2} l^* \exp(S(l^*)) \propto \\
&\propto \left( \frac{T}{V_C}\right)^{1/2} l^{\ast 3/2} \exp\left( -4J_{\perp}/T\right)  \exp\left(\tilde{C} V_C l^{\ast}/T\right)
\end{align*}
Neglecting pre-exponential numerical coefficients an inversion of this relation to $l^{\ast
}(\nu)$ and hence to $\mu^{\ast}(\nu)$ in the leading dependence for $T\rightarrow0$ yields
\[
\tilde{C} l^* \approx \frac{J_{\perp}}{V_C}+\frac{T}{8 V_C}\ln\left(  \frac{\nu^{2}V_C^{4}%
}{TJ_{\perp}^{3}}\right),\ \mu=2V_C\ln\frac{J_{\perp}}{V}
\]
It gives the saturated value of the rods length $l^{\ast}(T\rightarrow
0) \simeq J_{\perp}/V_C$ which is finite but large by our definition of the intermediate CI. The chemical potential saturates at the positive value.
This behavior is different from the one in the neutral system in both $D=2$ and $D=3$ dimensions. If we now try to find
$V_{inter}= J_{\perp}/\ln l^*$, we get an equation $\ln (\tilde{C} \cdot J_{\perp}/V_{inter}) = J_{\perp}/V_{inter}$ which possesses only
solutions $J_{\perp}/V_{inter} \sim 1$. This means that if we use $l^*$ as a characteristic length, then there is no intermediate CI regime in $2D$
in the unscreened case.

If the screening is more efficient, such that $l_{s}<l^*$, then $\ln l$
saturates to $\ln l_{s}$ and the CIs just shift the chemical potential as
$\mu^*\Rightarrow\tilde{\mu}=\mu^*-2V_C\ln l_{s}$. Then we get from (\ref{originalH}), (\ref{AppA_Jpar2D}), and (\ref{nu(mu)})
\begin{align}
\nu &  \approx \frac{1}{2}\exp\left( -\frac{4J_{\perp}}{T}\right) \left(  \frac{T}{\tilde{\mu}}\right)^{2},~
\tilde{\mu} \approx -\frac{T}{\sqrt{2\nu}}\exp\left( -\frac{2J_{\perp}}{T}\right)<0, \nonumber\\
~l^* & \approx\frac{T}{2|\tilde{\mu}|} \approx\sqrt{\frac{\nu}{2}}\exp\left( \frac{2J_{\perp}}{T}\right) > l_s \label{screened},
\end{align}
which coincides with the appropriate limit of the exact
result from \cite{Bohr:1983} under the shift $\mu\Rightarrow\tilde{\mu}$. Inversion
of this relation gives the chemical potential $\mu^*(\nu,T)  $ of
individual bisolitons as it is dictated by the reservoir of growing rods. For
$T\rightarrow0$ it saturates at a value $\mu^*\rightarrow 2 V_C\ln l_s$ but
$l^{\ast}$ keeps growing exponentially in $1/T$. In this case the regime of intermediate CIs is determined by $l_s$: $J_{perp} \gg V \gtrsim
V_{inter} = J_{\perp}/\ln l_s$.

In the absence of external carriers, the screening length $l_s$ is defined
self-consistently from the contribution of bisolitons and their rods. We shall
do it assuming the $3D$ media composed by noninteracting, except sharing the
Coulomb potential, layers -- that is not the case of our modeling which took a
$2D$ layer embedded into the $3D$ space. By definition, the dimensional screening length $a_{\perp}l_{s}$ is
given by the derivative of the $3D$ charge density $e\nu/(a_{||}a_{\perp}^{2})$
over the chemical potential $d\nu/d\mu \approx 4\nu l^{\ast}/T$. The
last relation follows from differentiation of (\ref{screened}).
We get
\begin{align*}
\frac{1}{l_{s}^{2}}=\frac{4\pi(2e)^{2}}{\epsilon a_{||} a_{\perp}^2}\frac{d\nu/2}{d\mu} &\approx
32 \pi\nu\frac{V_C}{T}\frac{l^*}{a_{||} a_{\perp}} ~, \\
~\frac{(l^* a_{\perp})^2}{l_{s}^{2}} &\approx 32\pi\nu\frac{V_C}{T}\frac{a_{\perp}}{a_{||}} l^{\ast3}%
\end{align*}
We see that the condition of the nonscreened regime
$l^{\ast}<l_{s}$ can be satisfied only for very small $\nu$ and still at not
very low $T$. The most common case will be described by relations
(\ref{screened}), which imitates the neutral system but with the up-shifted
chemical potentials and the corresponding strong increase of concentration of
noncondensed bikinks.

\subsection{$3D$ case}

Consider now the $3D$ case. Suppose the basic units are the straight bi-discs
(lenses) of unseparated bisolitons of radii $R$ (dimensionless; the physical
scale is $a_{\perp}$).

The CI energy of a disk is%
\[
\frac{4e^{2}}{\epsilon a_{\perp}^{4}}\int\int\frac{d^{2}rd^{2}r^{\prime}%
}{|{\mathbf r}-{\mathbf r}^{\prime}|}\exp\left(  -\frac{|{\mathbf r}-{\mathbf r}^{\prime}%
|}{a_{\perp}l_{s}}\right)  \approx 4\pi V_{3}R^{2}\min\{R,l_{s}\},
\]
where $V_{3} = C V_C = Ce^2/\epsilon a_{\perp}$ with $C=16/3$ for $R \ll l_s$ and $C=2\pi$ for $R \gg l_s$. The screening length $l_{s}$ appears for $R>l_{s}$.
Assumption of intermediate CIs (which are locally weak, but prevent the transverse walls formation) in the $3D$ case becomes $J_{\perp} \gg
V_3 \gg J_{\perp}/(H \min\{H,l_s\})$ (the cross-section of the sample is assumed to be $H\times H$ square).

The equilibrium relation for partial chemical potentials is held as before:
$\mu_{R}=2\pi R^{2}\mu$.
Then the distribution of disks is
\begin{equation}
n(R)=\exp(\beta(-4\pi RJ_{\perp}+2\pi R^{2}\mu-4\pi V_{3}R^{2}\min
\{R,l_{s}\})\label{n(R)3d},%
\end{equation}
where the first term is the confinement energy lost over the perimeter of the circle.
However new items appear in the $3D$ case, those are domain walls with the chemical potential $\mu_{wall} = \mu H^2$.
$\mu(\nu)$ is to be determined from the self-consistency condition
\begin{align}
&\nu = 2\sum_{R}\pi R^{2}n(l) + H^2 n_{wall} = \nonumber\\
& = 2\pi\sum_{R}R^{2}\exp(4\pi\beta(R^{2}\mu/2-J_{\perp} R-V_{3}R^{2}\min\{R,l_{s}\})) + \nonumber\\
&+ H^2 \exp(\beta\mu H^2 - \beta V_3 H^2 \min\{l_s, H\}).
\label{nu(mu)3D}%
\end{align}

Even in the case $V_3=0$, the wall term does not contribute to the sum as long as $\mu<0$. However, as $T$ decreases, $|\mu|$ also has
to decrease in order to accommodate all the solitons in the system. When $\mu$ reaches the value $\mu \propto -T/H^2 \approx 0$, the first
wall condenses, and then $\mu$ stays at $0$ analogously to Bose-Einstein condensation, but in the real space instead of the reciprocal one.

First, consider the unscreened case $l_{s}>H$. Then the summation in (\ref{nu(mu)3D}) is convergent
because of the two terms in the exponent: $J_{\perp}R$ as it was already
without CI, and now also $V_{3}R^{3}$ from the CI. This sum is dominated by
the minimal $R\varpropto1$ if either $\beta J_{\perp}\gg1$ or $\beta V_{3} \gg 1$.
Recall that our definition of a moderate CI means that it is not
efficient at the level of minimal distances, i.e. $J_{\perp}\gg V_{3}$, hence
only the first inequality is sufficient. Then
\[
\nu\approx n(1)\approx \exp(2\beta\mu - 8\beta J_{\perp}),
\]
and the estimation for the transverse aggregation temperature $T_{2}\varpropto J_{\perp}/\ln(1/\nu)$, obtained for the neutral case, still holds
here.

However, when the excess number of solitons condenses to first walls, for them, $R$
reaches the sample width $H$. Then for weak CIs $V_{3} \ll J_{\perp}/H^2$ when it is still unimportant
hence qualitatively never important at all. In this case $T_2$ only slightly decreases by $\Delta T_2 \propto V_3 H^2/\ln(1/\nu)$.

\begin{figure}[tbh]
\centering
\includegraphics[width=1.0\linewidth]{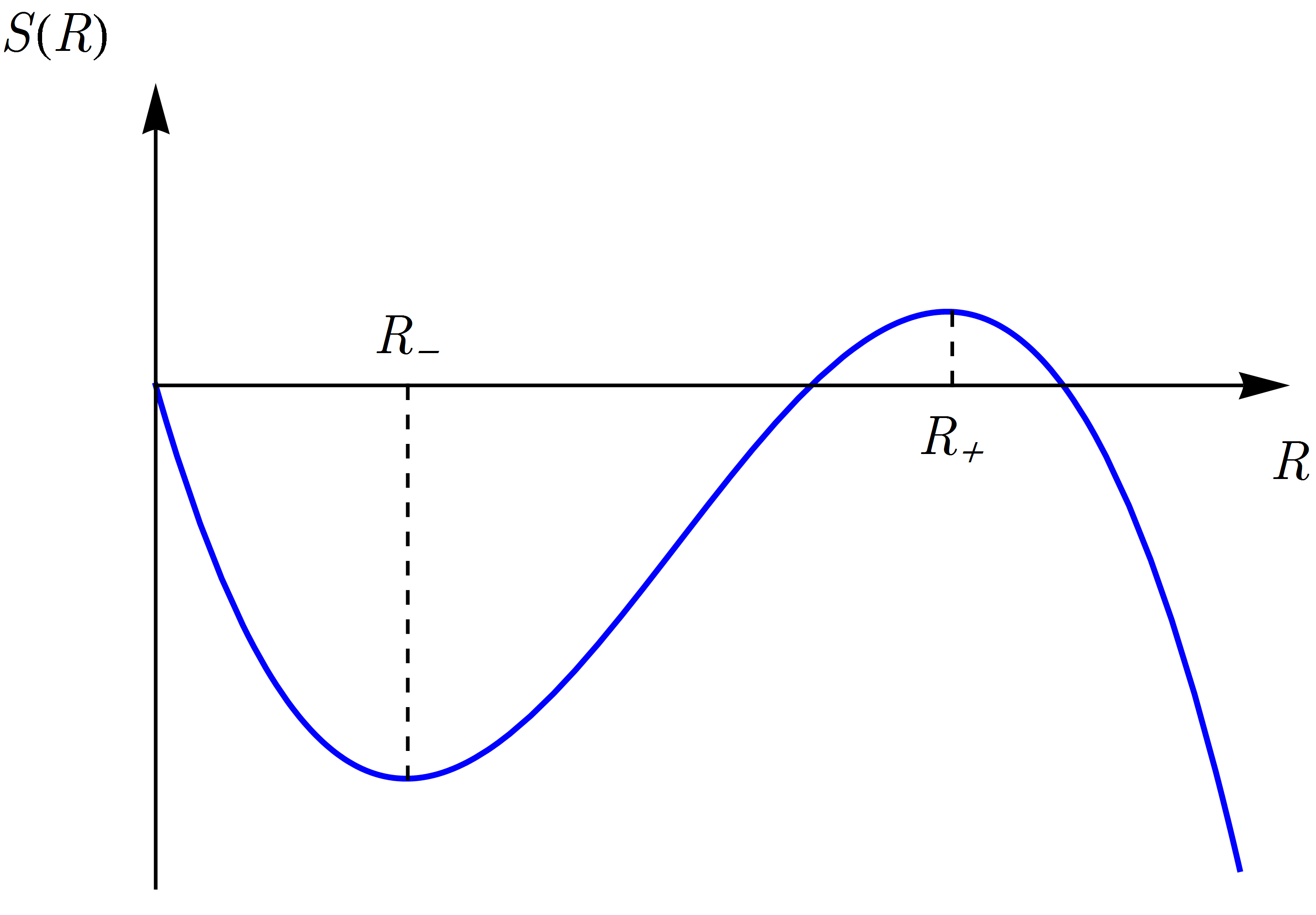}%
\caption{(Color online)  $S(R)$-dependence and its two local extrema $R_{\pm}$.}
\label{AppendixB-plot}
\end{figure}

For the moderate CI: $J_{\perp} \gg V_{3} \gg J_{\perp}/H^2$ walls do not form.
We can estimate the maximum size of disks analogously to the $2D$ case. When $T$ decreases, the chemical potential $\mu<0$ grows, then it reaches
$\mu=0$ and continues to increase. This means that the sum (\ref{nu(mu)3D}) is now determined by large values of $R$ and summation can be
approximated by integration over $R$,
\begin{equation}
\nu = 2\pi \int dR ~ R^2 \exp(4\pi\beta(R^{2}\mu/2-J_{\perp}R-V_{3}R^3)).
\label{int-nu-3D}
\end{equation}
The exponent $S(R)=4\pi\beta(R^{2}\mu/2-J_{\perp}R-V_{3}R^3)$ in (\ref{int-nu-3D}) decreases even faster at large $R$ in comparison to the $2D$
case, and the integral converges. $S(R)$ has local extrema at points $R_{\pm} = (\mu \pm \sqrt{\mu^2-12 V_3 J_{\perp}})/V_3$
(Fig.\ref{AppendixB-plot}). At some critical value $\mu \geq \mu_{cr} = 4\sqrt{V_3 J_{\perp}}$ the local maximum becomes positive: $S(R_+) \geq
0$, which means that large disks of radius $R^* = R_+(\mu_{cr}) = \sqrt{J/V_3} \gg 1$ appear in the system. In order to find the corresponding
critical $\nu_{cr}(T) = \nu(\mu_{cr},T)$ we use the saddle-point approximation for the integral over $R$ (only region $R\sim R^*$ contributes in
this case)%
\[
\nu_{cr}(T) \approx 2\pi R^{*2} \left( \frac{2\pi}{|d^{2}S/dR^{2}|}\right)^{1/2} e^{S(R^*)} \approx \pi T^{1/2} J^{3/4} V_3^{-5/4}
\]
At $T\rightarrow 0$, $\nu_{cr}(T)$ decreases below the fixed concentration $\nu$, which happens at $T\approx \nu^2 V_3 (V_3/J)^{3/2}$, then the
large disks appear.

Now turn to the screened case. For $R^*< l_s < H$, the screening is not very important: it does not affect even the largest disks' structure,
but only affects the interactions among them. However, for $l_s < R^*$, the screening becomes essential, since disks can grow up to $R^*$ and CIs does not prevent their further growth, so they grow into domain walls. The wall term appears in the sum (\ref{nu(mu)3D}) when the chemical potential
reaches the value $\mu \approx V_3 l_s$.
So, as in the $2D$ case, at $l_s < R^*$ there is just the shift of the chemical potential $\tilde{\mu} = \mu - V_3 l_s$.

In conclusion, the very weak CI $V_3 \ll J_{\perp}/H^2$ only shifts down the transition temperature $T_2$ not preventing the formation of the walls.
For the intermediate CI $J_{\perp}/H^2 \ll V_3 \ll J_{\perp}$, different regimes are observed depending on the screening length $l_s$. For $l_s >
R^*$, the CI is effectively unscreened, so it prevents the wall formation and only disks of maximum size of $R^* \approx \sqrt{J_{\perp}/V_3}
\gg 1$ are observed. For $l_s < R^*$, the screened CI does not prevent them growing to transverse domain walls and only shifts the chemical
potential.

\end{document}